\documentclass[journal]{IEEEtran}
\usepackage{amsfonts}
\usepackage{graphicx}
\usepackage{algorithm}
\usepackage{algorithmic}
\usepackage{amssymb}
\usepackage{amsmath}
\usepackage{setspace}
\usepackage{xcolor}
\usepackage{graphicx}
\usepackage{lipsum}
\usepackage{mathtools}
\usepackage{textcomp}
\usepackage{graphicx}
\usepackage[caption=false]{subfig}
\usepackage{graphicx}
\usepackage{epstopdf}
\usepackage{color}

\usepackage[acronym]{glossaries}
\makeglossaries
\newacronym{ekf}{EKF}{extended kalman filter}
\newacronym{cdrl}{CDRL}{constrained deep reinforcement learning}
\newacronym{drl}{DRL}{deep reinforcement learning}

\begin{document}
\title{Constrained Deep Reinforcement Learning for Cognitive Radar Resource Management}

\author{\IEEEauthorblockN{Ziyang Lu,  M. Cenk Gursoy,  Chilukuri K. Mohan,  Pramod K. Varshney} 

\thanks{The authors are with the Department of Electrical Engineering and Computer Science, Syracuse University, Syracuse NY, 13066. 

Email: \{zlu112, mcgursoy, ckmohan, varshney\}@syr.edu.
}

\thanks{The work in this paper is supported in part by the National Science Foundation Grant CNS 2221875.}
}


\maketitle

\begin{abstract}
In this paper, multi-target tracking and scanning are considered in a radar system operating in the track-while-scan mode. Specifically, time allocation for radar scanning and tracking of multiple maneuvering targets under a time budget constraint is addressed, aiming to jointly optimize the performance of both tracking and scanning in a cognitive radar. We first present the details of the model for tracking and scanning and formulate the time management task as a constrained optimization problem. Subsequently, we design a \gls{cdrl} framework to find the time allocation strategy for the problem. In the proposed \gls{cdrl} framework, the parameters of the neural networks and the dual variable are learned simultaneously. The deep deterministic policy gradient (DDPG) algorithm is introduced to tackle continuous action space and its performance is compared with deep Q-learning, heuristic approaches, and an optimization-based approach. Numerical results show that the radar with the proposed \gls{cdrl} framework can autonomously allocate more time to the tracking task that requires greater attention while providing time for scanning and also constraining the total time budget below the predefined threshold.
\end{abstract}

\begin{IEEEkeywords}
Constrained optimization, extended Kalman filter, multi-target tracking and scanning, track-while-scan, cognitive radar, reinforcement learning, resource allocation.
\end{IEEEkeywords}

\section{Introduction}
\subsection{Background}

Cognitive radar is a radar system that leverages advanced signal processing techniques as well as machine learning to optimize its performance in dynamic environments. Unlike traditional radar systems that operate with fixed parameters, cognitive radar systems continuously sense the environment and adapt their operational strategies in real-time.

Resource allocation in a cognitive radar is a critical aspect of its functionality. The radar system must efficiently allocate its limited resources, such as time, energy, and bandwidth to various tasks including scanning for newly emerging targets and tracking previously detected targets. Effective resource allocation ensures that the radar maintains a balance between these competing tasks, optimizing its performance without exceeding the resource budget constraints. The challenge lies in dynamically adjusting the resource allocation strategy to maximize the radar's effectiveness in a resource-constrained scenario.

\subsection{Prior Work}
In \cite{charlish2020development}, an overview of cognitive radar concepts is presented, highlighting three hierarchical levels of cognitive radar architecture: extracting information, managing resources, and refining environmental knowledge. The authors also survey the evolution of radar resource management techniques, demonstrating a progression of increasing adaptivity that culminates in the concept of cognitive radar.

Indeed, efficient allocation of radar resources is critical and has been extensively studied in the literature. For instance, radar resource management has been addressed through conventional optimization approaches in previous studies such as \cite{orman1996scheduling} and \cite{butler1997resource}. Another line of research employs game theoretical approaches. For instance, the authors in \cite{deligiannis2017game} formulate the power allocation problem in a multi-radar system as a non-cooperative game and perform an analysis of the Nash equilibrium and its convergence. 


Several prior works have addressed radar resource management for multi-target tracking and scanning. For example, the work in \cite{5438453} presents a hierarchical framework for task and dwell-level allocation, leveraging multi-criteria decision tools and operator-defined priorities to support complex radar scheduling. While effective, this approach relies on structured scheduling policies and does not adapt autonomously to changing environments. More recently, the work in \cite{10618738} investigates joint power and time allocation in cooperative multi-radar networks using a structured optimization framework. This method demonstrates strong performance under ideal conditions but assumes full knowledge of system parameters and performance models, limiting its flexibility in real-world scenarios.

The method proposed in \cite{7266661} formulates radar scheduling as a convex optimization problem, balancing search and tracking tasks by minimizing a weighted sum of detection and tracking costs. This approach achieves efficient allocation but operates myopically on a per-scan basis and requires known target arrival characteristics and environmental structure.

The work in \cite{schope2019optimal} proposes a radar resource management (RRM) framework for multi-target tracking, where dwell time allocation is optimized using a constrained Lagrangian relaxation approach. With the same RRM framework, a subsequent work in \cite{schope2022optimal} extends this problem to optimization for joint tracking and classification tasks, and an approximately optimal dynamic budget balancing (AODB) policy is derived using policy rollout in combination with Lagrangian relaxation.

Different from these prior works, our proposed \gls{cdrl} framework learns adaptive dwell time allocation policies through deep reinforcement learning, without requiring full observability or pre-defined scheduling rules. It operates in a single-radar system under partial observability of the environment, integrates $M$-of-$N$ track initiation logic, and provides a scalable and flexible approach to managing the trade-off between scanning and tracking in dynamic scenarios.

Recently, \gls{drl} has been extensively utilized to tackle dynamic decision-making problems, demonstrating high performance levels. Experimental results demonstrate that \gls{drl} surpasses previous algorithms in game playing and performs comparably to a human professional tester \cite{DBLP:journals/corr/abs-1207-4708}. As \gls{drl} algorithms become faster and more stable, they have been extensively researched in various fields, including resource allocation in cognitive radar systems.

For instance, in \cite{thornton2020deep}, the radar system dynamically selects frequency bands based on the locations and velocities of the targets, thereby avoiding collision with the coexisting communication system and sufficiently utilizing the available bandwidth. In \cite{stephan2022scene}, \gls{drl} is employed to dynamically tune the tracking parameters based on the scene, improving multi-target tracking performance. The work in \cite{durst2021quality} presents a novel approach to address radar resource management with an Advanced Actor-Critic (A2C) network, improving the utility and tracking accuracy. Different from the work in \cite{durst2021quality}, our approach explicitly enforces resource constraints using a \gls{cdrl} framework, which combines the methods of Lagrangian relaxation and DRL. Additionally, the study in \cite{durst2021quality} restricts the problem in experimental results to tracking tasks while we address joint tracking and scanning. Other learning-based applications for radar can be seen in \cite{selvi2020reinforcement} and \cite{meng2021deep}. To the best of our knowledge, although there are numerous studies that have explored various aspects of cognitive radar systems and time allocation strategies, no prior research or similar work has utilized \gls{drl} for time allocation in the context of a cognitive radar system under a time budget constraint.

In this work, we formulate radar scanning and target tracking under budget constraints as a constrained Markov decision process (CMDP). Related learning-based studies on CMDP are presented in \cite{tessler2018reward}, \cite{schulman2015trust}, \cite{pham2018optlayer}, \cite{le2019batch} and \cite{fioretto2020lagrangian}. Specifically in \cite{tessler2018reward}, the authors proposed a multi-timescale algorithm to address the constrained optimization problem. In \cite{khairy2020constrained}, a learning-based decision-making algorithm is proposed for UAVs, optimizing the network utility while taking into account the energy constraints. In \cite{eisen2019learning}, the authors proposed a learning-based resource allocation approach in wireless networks under energy constraints. In our previous work \cite{Ziyang-INFOCOM24}, we proposed a learning-based time allocation framework for cognitive radars in an integrated sensing and communication (ISAC) scenario. In the radar literature, the application of \gls{drl} to the CMDP model has not yet been extensively studied.

\subsection{Contributions}

In this work, we address the time allocation problem in a multi-function cognitive radar system operating in the track-while-scan mode. The key contributions of this work are as follows:

\begin{itemize}
    \item We define a utility function that addresses the trade-offs between scanning for potential targets and tracking already detected targets. We propose a flexible framework that can weigh the importance of scanning vs. tracking in a given application scenario and enables the prioritization of tracked targets. To the best of our knowledge, this form of utility function for multi-objective optimization has not been extensively explored in the context of radar resource management using deep reinforcement learning.
    
    \item We formulate the considered problem as a constrained optimization problem and propose a \gls{cdrl} framework to address it, where the parameters of the neural networks and the dual variable are updated simultaneously.
    
    \item We further improve the proposed \gls{cdrl} framework, leveraging the DDPG algorithm. The incorporation of DDPG equips our framework with the capability to manage continuous action spaces efficiently. Moreover, this adoption of DDPG enables centralized control while preventing action space explosion that would occur with deep Q-learning (DQL). For example, with DQL, the action space grows exponentially with the number of targets when using discretized actions, whereas DDPG requires only linear growth in action dimensions through its continuous action approach.

    \item We provide numerical results that show that the designed \gls{cdrl} framework can learn an efficient time allocation strategy compared to the heuristic time allocation approach, complying with the predefined time budget constraint.
\end{itemize}

We further identify key differences from our previous studies in \cite{Ziyang-TCCN-23} and \cite{Ziyang-ICC24}. Compared to \cite{Ziyang-TCCN-23}, which only addressed target tracking with online-trained deep Q-learning, this work considers both scanning and tracking with pre-trained agents. This work also enhances our prior study \cite{Ziyang-ICC24} by introducing a more realistic cognitive radar environment, incorporating an $M$-of-$N$ track initiation model, and using the number of missed targets as an effective scanning performance metric. We also evaluate the framework in diverse scenarios, including varying target priorities and beam misalignment effects. In addition to the DQL-based approach in \cite{Ziyang-ICC24}, we propose a centralized DDPG-based \gls{cdrl} framework that handles continuous action spaces and jointly allocates time across all targets, improving scalability and decision efficiency. We further compare our approach against an optimization-inspired baseline to better assess the effectiveness of our proposed method.

For ease of reference, we provide below a list of acronyms and list of notations used throughout the paper in Tables \ref{tab:acronyms}  and \ref{tab:notation1}, respectively.

\begin{table}[htbp]
\small
\caption{List of Acronyms}
\label{tab:acronyms}
\centering
\begin{tabular}{|c|p{5.5cm}|}
\hline
\textbf{Acronym} & \textbf{Full Form} \\
\hline
DRL & Deep Reinforcement Learning \\
\hline
CDRL & Constrained Deep Reinforcement Learning \\
\hline
MDP & Markov Decision Process \\
\hline
DQN & Deep Q-Network \\
\hline
DQL & Deep Q-learning \\
\hline
DDPG & Deep Deterministic Policy Gradient \\
\hline
SNR & Signal-to-Noise Ratio \\
\hline
EKF & Extended Kalman Filter \\
\hline
UCA &  Uniform Circular Array \\
\hline
GNN &  Global Nearest Neighbor \\
\hline
EA &  Equal Allocation \\
\hline
DA &  Distance-based Allocation \\
\hline
OA &  Optimization-based Allocation \\
\hline
\end{tabular}

\end{table}

\begin{table}[htbp]
\small
\caption{List of Notations}
\label{tab:notation1}
\centering
\begin{tabular}{|c|p{6cm}|}
\hline
\textbf{Symbol} & \textbf{Description} \\
\hline
$\mathbf{x}_t$ & Target state vector at time $t$: $[x_t, y_t, \dot{x}_t, \dot{y}_t]^T$ \\
\hline
$\mathbf{F}_t$ & State transition matrix \\
\hline
$T_0$ & Radar revisit interval \\
\hline
$\mathbf{w}_t$ & Maneuverability noise vector \\
\hline
$\sigma_w^2$ & Maneuverability noise variance \\
\hline
$\mathbf{Q}_t$ & Maneuverability noise covariance matrix \\
\hline
$\mathbf{z}_t$ & Measurement vector: $[r_t, \theta_t]^T$ \\
\hline
$r_t$ & Measured range from radar to target \\
\hline
$\theta_t$ & Measured azimuth angle \\
\hline
$\mathbf{v}_t$ & Measurement noise \\
\hline
$\sigma_r^2, \sigma_\theta^2$ & Measurement noise variances (range, azimuth angle) \\
\hline
$h(\cdot)$ & Nonlinear measurement function \\
\hline
$\mathbf{H}_t$ & Jacobian of measurement function \\
\hline
$\mathbf{R}_t$ & Measurement noise covariance matrix \\
\hline
$\mathbf{\hat{x}}_{t|t-1}$ & Prior state estimate \\
\hline
$\mathbf{\hat{P}}_{t|t-1}$ & Prior covariance matrix \\
\hline
$\mathbf{x}_{t|t}$ & Updated (posterior) state estimate \\
\hline
$\mathbf{P}_{t|t}$ & Updated (posterior) covariance matrix \\
\hline
$\mathbf{K}_t$ & Kalman gain \\
\hline
$\tau_t^n$ & Dwell time allocated to target $n$ at time $t$ \\
\hline
$\tau_s$ & Time allocated for scanning \\
\hline
$\phi$ & Phase delay between radar beams \\
\hline
$\tau_{\text{beam}}$ & Time duration per radar beam \\
\hline
$c_t^n$ & Tracking cost of target $n$ at time $t$ \\
\hline
$N$ & Maximum number of tracked targets \\
\hline
$N_t^{\text{miss}}$ & Number of missed (undetected) targets at time $t$ \\
\hline
$w_n$ & Priority of tracking target $n$ \\
\hline
$\beta$ & Trade-off coefficient between scanning and tracking \\
\hline
$U_t$ & Utility function  \\
\hline
$\lambda_t$ & Dual variable for constraint optimization \\
\hline
$\Theta_{\max}$ & Maximum fraction of time allowed for tracking \\
\hline
$a_t$ & Action taken by agent (dwell time decisions) \\
\hline
$s_t$ & State of the DRL agent at time $t$ \\
\hline
$\gamma$ & Discount factor in reinforcement learning \\
\hline
$Q(s,a)$ & Action-value function (Q-function) \\
\hline
$\alpha$ & Learning rate of the dual variable\\
\hline
$d_{r,t}^n, d_{a,t}^n$ & Distance of target $n$ to radar and to asset \\
\hline
$L_{bm}$ & Beam misalignment loss \\
\hline
$P_t$ & Transmit power of the radar \\
\hline
$G_t, G_r$ & Transmit and receive antenna gains \\
\hline
$k$ & Boltzmann's constant \\
\hline
$T_s$ & System temperature \\
\hline
\end{tabular}

\end{table}

\section{Radar Tracking Model}
\label{sectionRadar}

This section addresses the multi-target tracking problem in a two-dimensional space, employing an \gls{ekf}. We present the details of the tracking model and how to apply \gls{ekf} to estimate the states of the targets.

\subsection{Target Motion Model}

In time slot $t$, the current state of a target is described as $\mathbf{x}_t = [x_t, y_t, \dot{x}_t, \dot{y}_t]^T$, where $(x_t, y_t)$ denotes the current location of the target on the two-dimensional plane and $\dot{x}_t$, $\dot{y}_t$ are the current horizontal and vertical velocities of the target. Considering a constant velocity model within the revisit interval, the next state evolves from $\mathbf{x}_t$ to
\begin{equation}
\mathbf{x}_{t+1} = \mathbf{F}_t \mathbf{x}_t + \mathbf{w}_t,
\end{equation}
where $\mathbf{F}_t \in \mathbb{R}^{4\times4}$ is the transition matrix defined as
\begin{equation}
	\mathbf{F}_t = \begin{bmatrix}
		1&  0&  T_t&  0\\
		0&  1&  0&  T_t\\
		0&  0&  1&  0\\
		0&  0&  0&1  \\
	\end{bmatrix}
\end{equation}
where $T_t$ refers to the revisit interval at time $t$, which is determined by the radar system for tracking a specific target \cite{schope2021constrained}. Above, $\mathbf{w}_t$ denotes the maneuverability (acceleration) noise, which is a multivariate zero-mean Gaussian noise with the covariance matrix
\begin{equation}
	\mathbf{Q}_t = \begin{bmatrix}
		T_t^4/4&  0&  T_t^3/2&  0\\
		0&  T_t^4/4&  0&  T_t^3/2\\
		T_t^3/2&  0&  T_t^2&  0\\
		0&  T_t^3/2&  0&T_t^2  \\
	\end{bmatrix} \sigma_{w}^2
\end{equation}
where $\sigma^2_{w}$ is the maneuverability noise variance of the target at time $t$ \cite{schope2021constrained}.

\subsection{Measurement Model}

The radar acquires the noisy measurements of the current range $r$ and the azimuth angle $\theta$ of the target in order to estimate its location. Let us denote the measurement vector by $\mathbf{z}_t$ and denote the non-linear function that maps $\mathbf{x}_t$ to $\mathbf{z}_t$ by $h(\cdot)$. We then express the relationship between the state vector and the measurement vector as follows:
\begin{equation}
	\mathbf{z}_t = h(\mathbf{x}_t) + \mathbf{v}_t = \left[\sqrt{x_t^2+y_t^2}, \quad \text{tan}^{-1}\left(\frac{y_t}{x_t}\right)\right]^T + \mathbf{v}_t
\end{equation}
where $\mathbf{v}_t = [v_{r,t}, v_{\theta,t}]^T$ denotes the measurement noise vector, which consists of the range measurement noise ($v_{r,t}$) and the azimuth measurement noise ($v_{\theta,t}$) at time $t$. $v_{r,t}$ and $v_{\theta,t}$ are zero-mean Gaussian noise components with variances $\sigma_{r,t}^2$ and $\sigma_{\theta,t}^2$, respectively. 

For mathematical convenience, we assume the radar is positioned at the origin. This assumption is valid for a monostatic radar system, where the transmitter and receiver are co-located, and measurements $r$ and $\theta$ of the targets are inherently determined by the relative location of the target to the radar. Regardless of the actual placement of the radar, target positions can always be transformed into a radar-centric coordinate system, ensuring the generality of this assumption.

The estimated measurement noise variance of the echo radar signal can be calculated based on the signal-to-noise ratio ($\text{SNR}_t$) reflected from the target in time slot $t$. In this work, it is assumed that $\text{SNR}_t$ only depends on the dwell time $\tau_t$ of the radar system and the distance $r_t$ between the target and the radar. $\text{SNR}_t$ can be formulated as \cite{schope2021constrained}, \cite{koch1999adaptive}
\begin{equation}
	\text{SNR}_t(\tau_t, r_t) = \text{SNR}_0\left(\frac{\tau_t}{\tau_0}\right)\left(\frac{r_t}{r_0}\right)^{-4}
    \label{SNR}
\end{equation}
where $\text{SNR}_0$, $\tau_0$ and $r_0$ are the reference values of the SNR, dwell time and the distance from the target to the radar. Then, the variance of measurement noise can be obtained from the SNR according to \cite{meikle2008modern}
\begin{equation}
	\sigma_{\bullet,t}^2 = \frac{\sigma_{\bullet,0}^2}{\text{SNR}_t(\tau_t, r_t)}
	\label{sigma}
\end{equation}
where $\bullet \in (r, \theta)$. $\sigma_{\bullet,0}^2$ denotes the reference value of the corresponding measurement noise variance. It is worth noting that the variance of the measurement noise decreases when a longer dwell time is allocated to the target or when the target moves closer to the radar according to Equations (\ref{SNR}) and (\ref{sigma}).

Note also that the mapping function $h(\cdot)$ between the measurements and the states is non-linear and hence \gls{ekf} is employed in this work. When using \gls{ekf} (as will be detailed further in the next subsection), an observation matrix $\mathbf{H}_t \in \mathbb{R}^{2\times4}$ is introduced to linearize the relationship between $\mathbf{z}_t$ and $\mathbf{x}_t$. $\mathbf{H}_t$ is defined as the Jacobian of the measurement function $h(\cdot)$:
\begin{equation}
	\mathbf{H}_t = \frac{\partial h(\cdot)}{\partial \mathbf{x}}\vert_{\mathbf{x}_t} =
	\begin{bmatrix}
		
		\frac{x_t}{\sqrt{x_t^2+y_t^2}} &  \frac{y_t}{\sqrt{x_t^2+y_t^2}} &  0&  0\\
		\frac{-y_t}{x_t^2+y_t^2}&  \frac{x_t}{x_t^2+y_t^2}&  0&  0\\

	\end{bmatrix}.
\end{equation}

Considering independent measurements, the covariance matrix of the measurements is given by
\begin{equation}
	\mathbf{R}_t =
	\begin{bmatrix}
		\sigma_{r,t}^2  & 0\\
		0               & \sigma_{\theta, t}^2
	\end{bmatrix}.
\end{equation}

\subsection{Extended Kalman Filter} \label{subsec:Kalman}
Kalman filter is a well-known algorithm for estimating the state of a process, using a series of measurements over time \cite{welch1995introduction}. Extended Kalman filter is a non-linear version of Kalman filter, which can be employed for target tracking in radar systems with non-linear measurements.

The \gls{ekf} consists of two phases, namely predict and update. For a target-tracking problem, a simplified version of the procedure is as follows:
\subsubsection{Predict}
\begin{equation}
	\mathbf{\hat{x}}_{t|t-1} = \mathbf{F}_t \mathbf{x}_{t-1|t-1}
	\label{Kalman1}
\end{equation}
\begin{equation}
	\mathbf{\hat{P}}_{t|t-1} = \mathbf{F}_t \mathbf{P}_{t-1|t-1}\mathbf{F}_t^T + \mathbf{Q}_t
	\label{Kalman2}
\end{equation}
\subsubsection{Update}
\begin{equation}
	\mathbf{K}_{t} = \mathbf{\hat{P}}_{t|t-1}\mathbf{H}_t^T(\mathbf{H}_t \mathbf{\hat{P}}_{t|t-1} \mathbf{H}_t^T + \mathbf{R}_t)^{-1}
	\label{Kalman3}
\end{equation}
\begin{equation}
	\mathbf{x}_{t|t} = \mathbf{\hat{x}}_{t|t-1} + \mathbf{K}_{t}(\mathbf{z}_{t} - h(\mathbf{\hat{x}}_{t|t-1}))
	\label{Kalman4}
\end{equation}
\begin{equation}
	\mathbf{P}_{t|t} = (\mathbf{I} - \mathbf{K}_t\mathbf{H}_t) \mathbf{\hat{P}}_{t|t-1}
	\label{Kalman5}
\end{equation}
where $\mathbf{\hat{x}}_{t|t-1}$ and $\mathbf{\hat{P}}_{t|t-1}$ are the prior estimate and the corresponding covariance matrix at time $t$, respectively. $\mathbf{x}_{t|t}$ and $\mathbf{P}_{t|t}$ are the posterior estimate and the corresponding covariance matrix, respectively.

In the prediction phase, we use (\ref{Kalman1}) to compute the prior state estimate $\mathbf{\hat{x}}_{t|t-1}$ with its covariance matrix defined in (\ref{Kalman2}). In the update phase, the optimal Kalman gain is defined as in (\ref{Kalman3}), which is a weight given to the measurements and the current estimate. Eq. (\ref{Kalman4}) can be utilized to fuse the received measurements with the current estimate and obtain the posterior state estimate $\mathbf{x}_{t|t}$. The posterior covariance matrix $\mathbf{P}_{t|t}$ is computed via (\ref{Kalman5}).

In this work, the state vector $\mathbf{x}_{t|t}$ is initialized using available measurements from the track initiation process (introduced in Section \ref{subsec:initiation}), while $\mathbf{P}_{t|t}$ is initialized to reflect the uncertainty in these initial estimates. Specifically, when a track is confirmed through the detection of multiple sequential measurements, the initial position estimate is set to the most recent measurement coordinates. The initial velocity is estimated using the last two valid measurements in the measurement history, which may not be consecutive due to missed detections (e.g., in patterns such as VVXV, where V represents a valid measurement and X represents a missed detection).

The initial covariance matrix $\mathbf{P}_{t|t}$ is initialized empirically to reflect realistic uncertainty levels. We set the diagonal elements to be
\begin{equation}
\mathbf{P}_{0|0} = \text{diag}([10000, 10000, 100, 100])
\end{equation}
where the position variance (10,000 m$^2$) accounts for measurement noise and the velocity uncertainties (100 m$^2$/s$^2$) reflect the additional uncertainty introduced by the finite difference estimation method.

This measurement-based initialization provides the \gls{ekf} with a realistic starting point that leverages available information during the track initiation process. The \gls{ekf} then continuously updates the covariance matrix as new measurements are obtained, recursively refining $\mathbf{x}_{t|t}$ by minimizing the trace of the posterior estimate covariance matrix $\mathbf{P}_{t|t}$. This approach significantly reduces the convergence time, particularly beneficial in multi-target scenarios where rapid track establishment is crucial for maintaining situational awareness and resource allocation efficiency.

\subsection{Tracking Cost Function} \label{subsec:trackingcost}

We consider a cost function similar to the one defined in \cite{schope2021constrained}. Specifically, the tracking cost function at time $t$ is defined as
\begin{equation}
	c_t(T_t, \tau_t) = \text{trace}(\mathbf{E} \mathbf{P}_{t|t}\mathbf{E}^T)
	\label{cost}
\end{equation}
where
\begin{equation}
	\mathbf{E} =
	\begin{bmatrix}
		1&0&0&0\\
		0&1&0&0
	\end{bmatrix}.
\end{equation}

According to (\ref{cost}), the tracking cost is a function of both the revisit interval ($T_t$) and the dwell time ($\tau_t$) and can be interpreted as the sum of the posterior estimate variances of the target's current estimated $x$-axis and $y$-axis positions.

Multi-target tracking is considered in this problem, and tracking multiple targets requires extending the single-target model by running independent \glspl{ekf} for each target. A key challenge in this setting is data association, which ensures that each received measurement is correctly assigned to the corresponding target track. To address this, we use the global nearest neighbor (GNN) approach, where measurements are associated with the predicted target states based on the minimum distances.

In the multi-target setting, we denote the tracking cost for target $n$ at time $t$ as $c_t^n (T_t^n, \tau_t^n)$. Due to fixed revisit time $T_t^n = T_0$, the notation is further simplified to $c_t^n (\tau_t^n)$.

\section{Radar Scanning Model}

\subsection{Uniform Circular Array (UCA) Radar}
We consider a uniform circular array (UCA) radar system. It is assumed that the scanning beams are swept around the entire circular array in a phased-array manner, resulting in a 360\textdegree \, coverage of the scanning area. Note that radar scanning is performed in order to detect targets that may have emerged recently and have not been detected yet. 

A UCA radar can send scanning beams with a specific phase delay in between, which can steer the radar beam and uniformly sweep the entire circular scanning area. We formulate the total scanning time as
\begin{equation}
    \tau_{s} = \frac{360^\circ}{\phi}\tau_{beam}
    \label{beam_duration}
\end{equation}
where 
$\phi$ is the phase delay between adjacent radar beams, and $\tau_{beam}$ is the time duration of each beam. Although the scanning model is considered to be continuous, the radar transmits beams in a finite number of directions separated by an angular increment $\phi$ in the simulation. Thus, $\phi$ directly determines the angular resolution of the scanning process: smaller values of $\phi$ yield finer angular granularity (narrower beams), while larger $\phi$ leads to coarser scanning with fewer beams covering the 360$^\circ$ field of view.

\subsection{Measurement Acquisition Model During Scanning} \label{subsec:measurement}
In this work, the tracking dwell times $\{\tau_t^n\}_{n=1}^N$ (for $N$ targets) are determined by a DRL agent and the remaining time within the revisit interval is allocated to scanning the environment for detecting newly emerging targets. Then the time duration of each beam can be calculated according to (\ref{beam_duration}). The received SNR at the radar receiver during the scanning phase is denoted by $\text{SNR}_{\text{scan}}$ and can be derived as
\begin{equation}
    \text{SNR}_{\text{scan}} 
    = \frac{P_t \tau_{beam} G_t G_r \lambda^2 \sigma}{(4\pi)^3r^4LkT_s} \label{eq:SNRscan}
\end{equation}
where $P_t$ is the transmit power, $G_t$ and $G_r$ are the transmit and receive antenna gains, $\lambda$ is the radar signal wavelength, $\sigma$ is the radar cross section of the target, $r$ is the radar-target distance, $L$ is a loss factor, $k$ is Boltzmann's constant, and $T_s$ is the system temperature.

It is further assumed that the probability of a false alarm is fixed as $P_f$. During the simulation, we utilize Shnidman’s equation \cite{shnidman2002determination} to efficiently find the probability of detection $P_d$ given the received SNR and the probability of false alarm $P_f$. Instead of computing $P_d$ in real-time, we precompute a lookup table mapping SNR to $P_d$ over a range of probability of detection values from 0.01 to 0.99 in increments of 0.01. 

During operation, given an instantaneous value of the received SNR, the corresponding $P_d$ is efficiently determined using a nearest-neighbor search in the lookup table. With this characterization, a target present in the environment will be detected with probability $P_d$ that depends on its location.

As indicated by (\ref{eq:SNRscan}),  $\text{SNR}_{\text{scan}}$ in the scanning phase is influenced by the duration $\tau_{beam}$ of each beam and the distance $r$ between the target and the radar. The variance of the measurement error is inversely proportional to $\text{SNR}_{\text{scan}}$, similar to that in (\ref{sigma}). Hence, the detection probability $P_d$ will vary depending on $\tau_{beam}$ and $r$.

We further note that in each time slot, the radar has a probability $P_f$ of receiving a measurement due to a false alarm. It is assumed that the measurements resulting from false alarms are uniformly distributed across the area of interest.

\subsection{M-of-N Track Initiation Model} \label{subsec:initiation}

We adopt the standard $M$-of-$N$ model for initiating target tracks. The 3-of-4 track initiation model is selected for demonstration without loss of generality. The proposed framework can accommodate other $M$-of-$N$ configurations. Specifically, a track is established if $M = 3$ associated detections occur within $N = 4$ consecutive scans. Detection association is determined using the GNN algorithm based on a distance threshold $T_d$.

Each measurement buffer (up to $K$ total) stores the recent detection history using symbols: V indicates that an associated measurement was received at that time slot, and X indicates that no associated measurement was found. A track is initiated when three valid detections occur within four scans (e.g., VVV, VVXV, or VXVV). If this condition is not satisfied (e.g., VXX, VVXX), the buffer is reset.

A special case arises with the sequence VXVX: the older data is pruned, and the buffer transitions to VX, removing information no longer relevant to initiation.

Track initiation requests are handled using a first-come-first-served (FCFS) policy. When $K$ tracks are active, a new track can only be initiated once an existing one is terminated.

\subsection{Scanning Metric}

The metric employed to evaluate scanning performance is defined as the difference between the total number of targets present in the environment and the number of targets currently being tracked. Hence, the goal is to keep this difference (i.e., the number of undetected targets) small and ideally equal to zero to avoid any missed targets. Since the ground truth of the number of targets present in the environment is unknown, this metric is only used as part of the reward function during the training phase and for evaluation purposes. In practice, this scanning metric may also be affected by false tracks arising from clutter or false alarms. In our simulations, we set the false alarm probability to a low value ($P_f = 10^{-3}$) to minimize such occurrences.

\section{Problem Formulation}

\subsection{Objective Function}

In the previous sections, radar tracking and scanning and their corresponding performance metrics have been presented. Specifically, the tracking cost of target $n$ at time $t$ is $c_t^n (\tau_t^n)$ as given in (\ref{cost}), where $\tau_t^n$ denotes the dwell time allocated for tracking target $n$ in time slot $t$. For the tracking phase, the goal is to minimize the total weighted tracking cost for all the targets. On the other hand, for the scanning phase, the goal is to minimize the number of undetected targets that are currently not being tracked.

We propose the following utility function as a measure of the joint performance of the radar system:
\begin{equation}
    U_t(\{\tau_t^n\}_{n=1}^N) = -\sum_{n=1}^N w_n c_t^n (\tau_t^n) - \beta N_t^{miss},
    \label{utility}
\end{equation}
where $N$ is the number of targets the radar system is tracking, $w_n$ is the weight parameter indicating the priority associated with tracking target $n$,  $N_t^{miss}$ denotes the number of undetected targets that are not being tracked during time slot $t$, and $\beta > 0$ is the tradeoff coefficient introduced to address the performance tradeoff between tracking and scanning. The value of $\beta$ can vary depending on specific scenarios and system requirements. Setting $\beta$ too low may reduce its impact, while excessively high values of $\beta$ could either cause instability in the learning process or force the agent to focus on the scanning performance. Note that the utility function is always negative or zero since both the tracking costs and the number of missed targets are non-negative. This formulation enables us to frame the radar resource management problem as a maximization problem.

We consider a time-slotted system, where the duration of each time slot is $T_0$, which is the radar revisit interval that stays fixed across different time slots. Within each time slot, the strategic allocation of the available time $T_0$ between tracking and scanning tasks and also allocation among multiple tracked targets is critical. For instance, allocating more time to tracking can reduce the tracking cost for currently tracked targets (i.e.,  improve tracking accuracy), whereas dedicating more time to scanning can reduce the number of missed targets and the target initiation time. Additionally, effective allocation of time among the targets already being tracked is also crucial for minimizing the total tracking cost.  Therefore, our goal is to find an efficient time allocation policy to balance the performance of the tracking and scanning phases and optimize the overall performance of the radar system.

\subsection{Markov Decision Process (MDP) Formulation}

The radar resource allocation problem is formulated as a CMDP, which extends the standard MDP by incorporating resource constraints. A standard MDP is defined as a tuple:
\begin{equation}
    \mathcal{M} = (S, A, P, R, \gamma),
\end{equation}
where
\begin{itemize}
    \item \( S \) is the state space, representing the partial observations of the radar environment, including tracking costs, past time allocations, and the dual variable;
    \item \( A \) is the action space, and actions correspond to the dwell times allocated to targets;
    \item \( P(s' | s, a) \) represents the state transition probabilities, capturing how the radar environment evolves based on time allocation decisions;
    \item \( R(s, a) \) is the reward function, balancing tracking accuracy and scanning performance;
    \item \( \gamma \in (0,1] \) is the discount factor that weighs future rewards.
\end{itemize}

Unlike a standard MDP, which optimizes only the expected cumulative reward, our problem includes an additional time budget constraint, making it a CMDP. This constraint ensures that the total dwell time allocated for tracking does not exceed a predefined budget.

\subsection{Constrained Optimization Problem}

Following \cite{tessler2018reward} and \cite{altman1999constrained}, we can formulate the time allocation problem in the radar system as the following constrained sequential optimization problem:
\begin{equation}
	\begin{aligned}
		\max_{\pi}  \quad & \sum_{m=0}^\infty \Bigg[\gamma^m U_{t+m}(\{\tau_{t+m}^n\}_{n=1}^N)\Bigg]\\
		\textrm{s.t.} \quad & \sum_{m=0}^\infty\gamma^m\left(\sum_{n=1}^N \frac{\tau_{t+m}^n}{T_0} - \Theta_{\max}\right) \leq 0
	\end{aligned}
	\label{formulated_constrained_opt}
\end{equation}
where $U_t(\{\tau_t^n\}_{n=1}^N)$ is the utility function defined in (\ref{utility}), $\Theta_{\max} \in (0,1]$ is the time budget limit, and $\gamma \in (0,1)$ is the discount factor. Note that the (normalized) 
time budget allocated to tracking target $n$ is defined as the ratio of dwell time $\tau^n$ of target $n$ to radar revisit interval $T_0$, i.e., $\frac{\tau^n}{T_0}$. 

As shown in the optimization problem (\ref{formulated_constrained_opt}), the goal is to find a time budget allocation policy $\pi$ that maximizes the discounted sum utility function while satisfying the discounted sum budget constraint. In this work, we employ \gls{drl} to determine such a policy. More specifically, the policy $ \pi $ is a stochastic function that maps a given state $ s_t $ to an action $ a_t $ with a probability distribution, denoted as $ \pi(a_t | s_t) $, where the state $s_t$ is the partial observations of the radar environment and the action $a_t = \{\tau_{t}^n\}_{n=1}^N$ is the time allocation decisions at time $t$. In deep reinforcement learning, this policy is represented by neural networks. The objective of reinforcement learning is to optimize the neural network parameters to maximize the expected cumulative reward, enabling the policy to make optimal decisions over time.

Before we design the DRL agent, we note that by introducing a non-negative dual variable $\lambda_t$ (i.e., by utilizing Lagrangian relaxation), the problem in (\ref{formulated_constrained_opt}) can be relaxed to an unconstrained optimization problem as follows: 

\begin{align}
\tiny	
 \hspace{-.3cm}\min_{\lambda_t \geq 0} \max_{\pi} \sum_{m=0}^\infty \gamma^m \left[U_{t+m}(\{\tau_{t+m}^n\}_{n=1}^N)\!-\!\lambda_t\!\left(\sum_{n=1}^N \frac{\tau_{t+m}^n}{T_0} \!-\! \Theta_{\max}\!\right)\right]
	\label{dual}
\end{align}

To solve the optimization problem under this CMDP, we employ a \gls{cdrl} framework, which approximates the optimal policy using neural networks. The dual variable \( \lambda_t \) is learned alongside the policy using a gradient-based update. Details will be presented in the next section.

\section{Constrained Deep Reinforcement Learning Framework with Deep Q-learning}
\label{sec_dql}

\subsection{Deep Q-Learning}

Deep Q-learning (DQL) is a reinforcement learning algorithm that combines Q-learning with deep neural networks to handle high-dimensional state spaces. The goal of DQL is to find a policy that maximizes the expected discounted sum of future rewards:

\begin{equation}
    R_t = \sum_{m=0}^{\infty} \gamma^m r_{t+m},
\end{equation}
where \( \gamma \in (0,1] \) is the discount factor that determines the relative importance of future rewards, and \( r_{t} \) is the instantaneous reward at time \( t \).

Let \( Q^\pi(s, a) \) denote the action-value function (Q-function), which represents the expected discounted sum of rewards when taking action \( a \) in state \( s \), following policy \( \pi \):

\begin{equation}
    Q^\pi(s, a) = \mathbb{E} \left[ \sum_{m=0}^{\infty} \gamma^m r_{t+m} \,\bigg|\, s_t = s, a_t = a \right].
\end{equation}

Here
\begin{itemize}
    \item \( s \) is the state of the radar system at time \( t \);
    \item \( a \) is the action, corresponding to the dwell time allocated for tracking;
    \item \( r_t \) is the immediate reward function;
    \item \( \gamma \) is the discount factor.
\end{itemize}

The goal of the learning agent is to estimate the optimal action-value function
\begin{equation}
    Q^*(s, a) = \max_{\pi} Q^{\pi}(s, a),
\end{equation}
which corresponds to the highest possible expected return achievable from state \( s \) and action \( a \). The agent uses this to derive the optimal policy that maximizes cumulative rewards.

The Q-values are updated iteratively using the Bellman equation
\begin{equation}
    \begin{aligned}
        Q_k(s, a) = Q_{k-1}(s, a) + \alpha \Big( r + \gamma \max_{a'} Q_{k-1}(s', a')  \\
        - Q_{k-1}(s, a) \Big),
    \end{aligned}
\end{equation}
where
\begin{itemize}
    \item $k$ represents the iteration index;
    \item $\alpha$ is the learning rate;
    \item $s'$ is the next state after taking action $a$ at state $s$;
    \item $a'$ is the action that maximizes the Q-value in state $ s'$.
\end{itemize}

This ensures that Q-values are progressively refined over iterations as the agent learns from interactions with the environment.

DQL employs a deep Q-network (DQN) to approximate the Q-values. The DQN is trained using experience replay and backpropagation. The agent stores its experience tuple \( (s, a, r, s') \) in a replay buffer, which is used to update the neural network in a more stable manner.

To balance exploration and exploitation, the agent follows the \( \epsilon \)-greedy strategy. The agent selects the action with the highest Q-value
\begin{equation}
    a_t = \arg\max_{a} Q(s, a),
\end{equation}
with probability \( 1 - \epsilon \), and selects a random action with probability \( \epsilon \) to encourage exploration. Over time, \( \epsilon \) is gradually reduced to favor exploitation of the learned policy.

\subsection{DQL-based \gls{cdrl} Framework}

We assume that resource management for the tracking of each target is performed by a DQL agent, and all DQL agents share the same copy of the DQN.

\subsubsection{State} State of agent $n$ at time $t$, denoted as $s_t^n$, consists of the partial observations of the current environment. More specifically, we define the state as follows:
\begin{equation}
    \mathbf{s}_t^n = \Bigg[w_n c_{t-1}^n, \{w_nc_{t-1}^n(\tau_{t-1}^n)\}_{n=1}^N,\{\tau_{t-1}^n\}_{n=1}^N,\lambda_{t-1}\Bigg].
    \label{state}
\end{equation}

The first entry $w_n c_{t-1}^n$ is the weighted tracking cost of target $n$ in the previous time slot. The next $N$ entries $\{w_nc_{t-1}^n(\tau_{t-1}^n)\}_{n=1}^N$ form the list of the weighted tracking costs of all the targets in time slot $(t-1)$. In scenarios in which the actual number of targets is fewer than $N$, the corresponding entries of this list are populated with zeros. In the above, $\{\tau_{t-1}^n\}_{n=1}^N$ represent the dwell times selected at time $(t-1)$, and $\lambda_{t-1}$ denotes the dual variable again in time slot $(t-1)$. Overall, the size of the state space is $2N + 2$.

\subsubsection{Action} DQL agent $n$ needs to select a dwell time for tracking its corresponding target. The action is defined as $a_t^n = \tau_t^n$, where the dwell time $\tau_{t}^n \in [0, T_0]$ represents the allocated duration (within the revisit interval of length $T_0$) dedicated to tracking target $n$ during time slot $t$. The available time is quantized into $L$ levels and hence the size of the action space is $L$. Once all the agents make their own decisions, the remaining time allocated to scanning can be computed as
\begin{equation}
    \tau_s = T_0 - \sum_{n=1}^N \tau_{t}^n.
\end{equation}

\subsection{Reward Function} To solve the unconstrained optimization problem in (\ref{dual}), the reward is defined as
\begin{equation}
    r_t = U_{t}(\{\tau_{t}^n\}_{n=1}^N)-\lambda_t\left(\sum_{n=1}^N \frac{\tau_{t}^n}{T_0} - \Theta_{\max}\right)
    \label{reward}
\end{equation}
where the first term on the right-hand side is the utility function and the second term is the penalty if the budget constraint is violated. The reward is shared among all $N$ DQL agents to enable cooperation among the agents. In the proposed \gls{cdrl} framework, the DQN parameters for maximizing the reward function and the value of the dual variable $\lambda_t$ are learned simultaneously.

The reward function is applied only during the training phase, where we assume the ground truth of the environment is known. This allows the DRL agent to receive well-defined feedback based on the tracking costs, the number of missed targets, and the budget constraint. The reward function serves as a guiding mechanism to optimize decision-making during training.

During the testing phase, the trained DRL agent is deployed as a state-to-action mapping function without requiring knowledge of the reward function. Since the agent has already learned an optimal policy, it can make decisions solely based on the observed state information. This ensures that the proposed framework remains practical for real-world deployment, where ground truth information is not available.

\subsection{Workflow of the DQL-based \gls{cdrl} Framework}
\begin{algorithm}[!ht]
	\caption{DQL-based \gls{cdrl} Algorithm}
	\begin{spacing}{1}
	\begin{algorithmic}[1]
		\STATE{Initialize the DQN parameters $\Phi_0^{\pi}$ with random values.}
		\STATE{Initialize states $\{\mathbf{s}_0^n\}_{n=1}^N$ as zero vectors and $\lambda_t$ as $\lambda_0$.}
		\FOR{time slot $t=0,1,..., T_{max}$}
		\FOR{each agent $n = 1, 2,..., N$}
		\STATE{Select an action $a_t^n$ based on the current state $\mathbf{s}_t^n$ with the DQN $\Phi_t^{\pi}$ and $\epsilon$-greedy method.}

		\ENDFOR
		\STATE{Compute reward $r_t$ according to (\ref{reward}).}
		\FOR{each agent $n = 1, 2,..., N$}
		\STATE{Store the experience ($\mathbf{s}_t^n$, $a_t^n$, $r_t$, $\mathbf{s}_{t+1}^n$) to the DQN experience buffer.}
		\STATE{Update $\Phi_t^{\pi}$ to $\Phi_{t+1}^{\pi}$ with experience replay and back-propagation.}
		\ENDFOR
		\STATE{$\lambda_{t+1} = \max(0,\; \lambda_t - \alpha (\sum_{n=1}^N \tau_{t}^n - T_0) )$.}
		\ENDFOR
	\end{algorithmic}
	\end{spacing}
	\label{algorithm1}
\end{algorithm}
The proposed DQL-based \gls{cdrl} algorithm is given in Algorithm \ref{algorithm1} above. We simultaneously update the DQN parameters $\Phi_t^{\pi}$ and the dual variable $\lambda_t$. The DQN is parameterized by $ \Phi_t^\pi $, which consists of the trainable parameters of the neural network.

The objective of the proposed \gls{cdrl} algorithm is to find a solution to the problem in (\ref{dual}). By defining the reward function as in (\ref{reward}), the DRL agent is trained to optimize the discounted reward $r_t$, i.e.
\begin{equation}
    \max_{\pi} \sum_{m=0}^\infty \gamma^m \left[U_{t+m}(\{\tau_{t+m}^n\})-\lambda_t\left(\sum_{n=1}^N \frac{\tau_{t+m}^n}{T_0} - \Theta_{\max}\right)\right].
    \label{DRLjob}
\end{equation}

We denote the objective function in (\ref{DRLjob}) as $\mathcal{L}$. Then, the dual variable $\lambda_t$ is updated by minimizing $\mathcal{L}$ over $\lambda_t$, i.e.,
\begin{equation}
    \begin{aligned}
	\lambda_{t+1} &= \max(0,\; \lambda_t - \alpha \bigtriangledown_{\lambda_t} \mathcal{L}) \\
	&= \max\left(0,\; \lambda_t + \alpha \sum_{m=0}^\infty \gamma^m \left(\sum_{n=1}^N \frac{\tau_{t+m}^n}{T_0} - \Theta_{\max}\right)\right)
	\end{aligned}
\end{equation}
where $\alpha$ is the learning rate of the dual variable. The gradient $\bigtriangledown_{\lambda_t} \mathcal{L}$ can be estimated with an additional neural network but we simplify the update of $\lambda_t$ instead as
\begin{equation}
	\lambda_{t+1} = \max\left(0,\; \lambda_t + \alpha \left(\sum_{n=1}^N \frac{\tau_{t}^n}{T_0} - \Theta_{\max}\right) \right).
\end{equation}

\begin{figure}[h!]
    \centering
    \includegraphics[width=.4\textwidth]{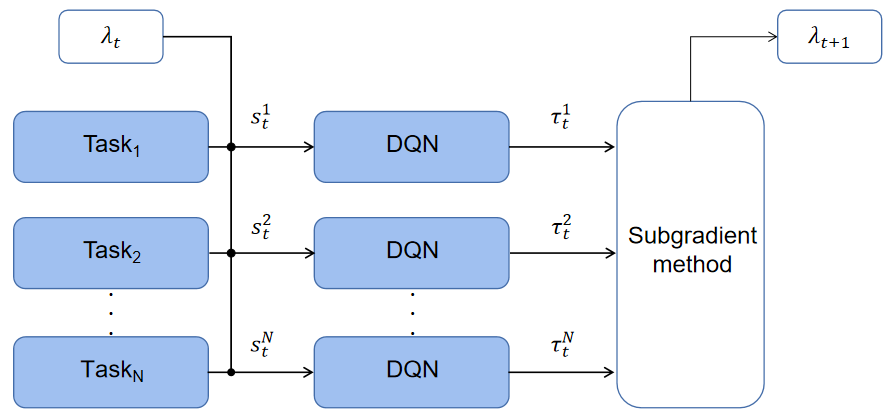}
    \caption{\gls{cdrl} framework with Deep Q-learning}
    \label{DQL}
\end{figure}

The workflow of the DQL-based \gls{cdrl} framework is depicted in Fig. \ref{DQL}. Intuitively, the dual variable $\lambda_t$ increases when the total time for tracking exceeds the predefined budget, thus increasing the penalty term in the reward function (\ref{reward}). Conversely, the dual variable decreases if the time constraint is not violated. The proposed \gls{cdrl} framework dynamically constrains the total budget below the budget constraint.

\section{Constrained Deep Reinforcement Learning Framework with Deep Deterministic Policy Gradient}

\subsection{Motivation}

In Section \ref{sec_dql}, we addressed the problem using the DQL algorithm. In this setting, each time DQN is utilized, a decision is made regarding the dwell time for a single target. Consequently, if a single DQN is used, each tracking task sequentially accesses the shared DQN to determine its respective dwell time. Equivalently, as illustrated in Fig. \ref{DQL}, a parallel configuration of DQNs can be utilized to simultaneously determine the dwell times for all targets.

To improve the time efficiency and computational complexity of the proposed framework, a single agent is preferred to make decisions for all tracking tasks simultaneously, adopting a centralized control approach. However, implementing centralized control with the DQL algorithm introduces significant challenges. In the DQL-based framework, each target-tracking task requires access to the DQN. This requires either (i) all agents \textit{taking turns to sequentially access the same DQN}, which leads to increased decision latency, or (ii) each agent \textit{maintaining a separate copy of the DQN}, which significantly increasing memory and computational overhead. Both approaches are inefficient for real-time radar resource management. 

Additionally, if a centralized DQL approach is adopted, the action space grows exponentially with the number of targets and the precision of dwell time quantization. Specifically, for \( N \) targets and a dwell time range discretized into \( L \) levels, the centralized action space has a size of \( L^N \), leading to an exponential increase in complexity as \( L \) or \( N \) grows. These limitations highlight the importance of using centralized control with DDPG, which inherently supports continuous action spaces and allows a single agent to efficiently determine dwell times for all targets in a single forward pass of the neural networks, thereby reducing computational overhead and improving decision speed.

\subsection{Deep Deterministic Policy Gradient (DDPG)}
To avoid the explosive growth in the action space, we consider continuous action spaces and develop a DDPG algorithm for radar resource management.

DDPG is an effective reinforcement learning algorithm specifically designed to address problems involving continuous action spaces \cite{lillicrap2015continuous}. The DDPG algorithm employs the actor-critic algorithm and consists of four neural networks: an actor network, a target actor network, a critic network, and a target critic network. The actor network and the target actor network have the same structure and they both take observations as  inputs and output deterministic continuous actions. Similarly, the critic network and target critic network have identical structures and they both take the observations and the action selected by the actor network as inputs, and output Q-values to evaluate the selected action. The presence of target networks improves the stability of the learning process. 

\subsection{DDPG-based \gls{cdrl} Framework}
Below, we describe the state and action spaces as well as the reward function of the proposed DDPG agent. We note that there is one DDPG agent deciding the dwell time for all the targets.
\subsubsection{State}
State $\mathbf{s}_t$ consists of the current partial observation of the environment. In particular, we define the state as follows:
\begin{equation}
    \mathbf{s}_t = \Bigg[\{w_nc_{t-1}^n(\tau_{t-1}^n)\}_{n=1}^N,\{\tau_{t-1}^n\}_{n=1}^N,\lambda_{t-1}\Bigg].
\end{equation}

The state structure is similar to that of the DQL-based framework as described in Section \ref{sec_dql}, except that the first term in the state is removed. The dimension of the state space is $2N+1$.

\subsubsection{Action and Reward}
In DDPG, the dwell times allocated to tracking all targets are determined simultaneously. Hence, the action is defined as $a_t = \{\tau_{t}^n\}_{n=1}^N$, where $\tau_{t}^n \in [0, T_0]$. Consequently, the dimension of the action vector is $N$. Similarly as before, the remaining time from the budget is allocated to radar scanning, i.e., $    \tau_s = T_0 - \sum_{n=1}^N \tau_{t}^n$.

Reward $r_t$ is the same as the setting in the DQL-based \gls{cdrl} framework, and hence is given by (\ref{reward}).

\subsection{Workflow of the DDPG-based \gls{cdrl} Framework}
\begin{figure}[h!]
    \centering
    \includegraphics[width=.4\textwidth]{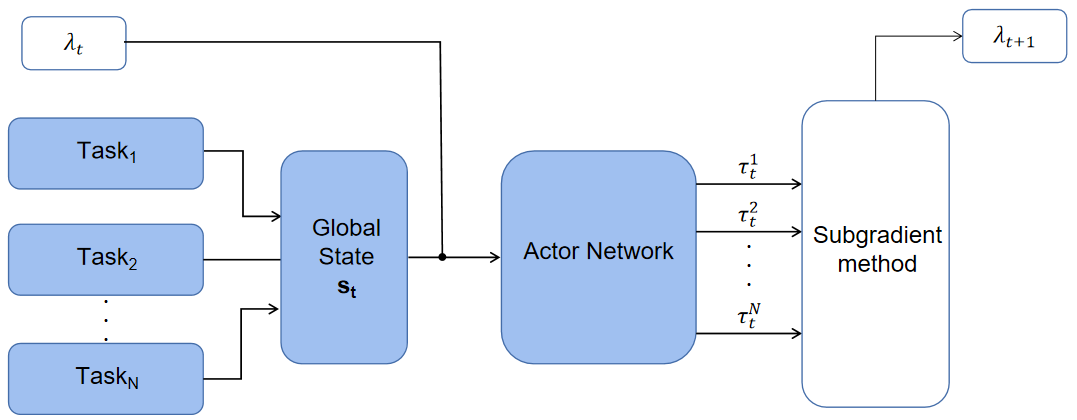}
    \caption{\gls{cdrl} framework with DDPG}
    \label{DDPG}
\end{figure}

\begin{algorithm}[!ht]

 \caption{DDPG-based \gls{cdrl} Algorithm}
	\begin{spacing}{1}
	\begin{algorithmic}[1]
		\STATE{Initialize the parameters of the DDPG networks with random values.}
		\STATE{Initialize state $\mathbf{s_0}$ as zero vectors and dual variable $\lambda_t$ as $\lambda_0$.}
		\FOR{time slot $t=0,1,..., T_{max}$}
		\STATE{Select action $a_t$ based on the current state              $\mathbf{s}_t$ with the actor network in the DDPG algorithm.}
  
		\STATE{Compute reward function $r_t$ according to                 (\ref{reward}).}
		
		\STATE{Store the experience ($\mathbf{s}_t$, $a_t$, $r_t$, $\mathbf{s}_{t+1}$) to the DDPG experience replay buffer.}
		\STATE{Update neural networks in DDPG with experience replay and back-propagation.}
		\STATE{$\lambda_{t+1} = \max(0,\; \lambda_t - \alpha (\sum_{n=1}^N \frac{\tau_{t}^n}{T_0} - \Theta_{max}) )$.}
		\ENDFOR
	\end{algorithmic}
	\end{spacing}
	\label{algorithm2}
\end{algorithm}
The DDPG-based \gls{cdrl} algorithm is presented in Algorithm \ref{algorithm2}. The primary difference between the DDPG-based framework and the DQL approach is the centralized control of the dwell time allocation with the single DDPG agent. The workflow of the DDPG-based framework can be seen in Fig. \ref{DDPG}.

\section{Numerical Results and Analysis}

\subsection{Experimental Setup}

\subsubsection{Hyperparameters}

\begin{table}[!ht]
	\caption{Simulation Parameters}
	\label{tab: environment}
	\centering
	\small
	\begin{tabular}{|c||c|}
 
        \hline \textbf{Simulation Parameters} & \textbf{Value} \\
		\hline Range measurement noise variance $\sigma_{r,0}^2$ ($m^2$)&16\\
		\hline Angle measurement noise variance $\sigma_{\theta,0}^2$ $(\text{rad}^2)$&1e-6\\
		\hline Maneuverability noise variance $\sigma_w^2$ ($(m^2/s^4)$) & 16\\
		\hline Reference distance $r_0$ ($m$) & 3000\\
		\hline Reference dwell time $\tau_0$ (s) & 1\\
		\hline Revisit interval $T_0$ (s) & 2.5\\
        \hline Tradeoff coefficient $\beta$ & 3e4 \\ 
        \hline Dwell Time budget $\Theta_{max}$ & 0.9\\
		\hline Initial dual variable $\lambda_0$ & 5000\\
		\hline Step size of dual variable $\alpha$ & 5000\\  
		\hline
		
	\end{tabular}
\end{table}

\begin{table}[htbp]
\caption{DQN Hyperparameters}
\label{tab:dqn_parameters}
\centering
\begin{tabular}{|c||c|}
\hline
\textbf{Hyperparameter} & \textbf{Value} \\
\hline
Learning rate & 0.0001 \\
\hline
Discount factor & 0.9 \\
\hline
Batch size & 32 \\
\hline
Replay buffer size & 1e6 \\
\hline
Exploration rate & 0.1 \\
\hline
\end{tabular}
\end{table}

\begin{table}[htbp]
\caption{DDPG Hyperparameters}
\label{tab:ddpg_parameters}
\centering
\begin{tabular}{|c||c|}
\hline
\textbf{Hyperparameter} & \textbf{Value} \\
\hline
Actor learning rate & 0.0002 \\
\hline
Critic learning rate & 0.0002 \\
\hline
Discount factor & 0.9 \\
\hline
Batch size & 128 \\
\hline
Replay buffer size & 1e6 \\
\hline
Soft update rate & 0.01 \\
\hline
\end{tabular}
\end{table}

The parameters of the simulation environment, the hyperparameters of the DQL agent and DDPG agent are provided in Tables \ref{tab: environment}, \ref{tab:dqn_parameters} and \ref{tab:ddpg_parameters}, respectively.

In the DQL agent, the DQN comprises two feedforward layers, each containing 64 neurons, with ReLU as the activation function between the layers.

In the DDPG algorithm, there are four networks. The actor network and its target network have identical structures, each with two layers containing 256 and 128 neurons, respectively, and ReLU activation functions between the layers.

Similarly, the critic network and its target network share the same structure, consisting of two layers with 100 neurons each, with ReLU as the activation function.

\subsubsection{Target Spawning Model}
In the experiments, the radar system is placed at the origin and it is assumed that a maximum of $N = 5$ targets can be tracked simultaneously. To enhance environmental diversity and evaluate the proposed framework's effectiveness under various conditions, we employ the following general target spawning model:

\begin{itemize}
    \item Every 100 iterations, a new target can join the environment with probability 0.03. The initial positions and velocities of the targets are generated randomly.
  
    \item To further diversify the operational environment, any target present for more than 3000 time slots is removed from the environment.
  
    \item It is assumed that the radar is monitoring an area with a radius of 20 km. Targets that move beyond this range are considered to have exited the environment and are no longer tracked by the radar.
\end{itemize}

    After training in the environment with the above target spawning model, the DQN-based and DDPG-based agents are tested in a new, unfamiliar environment. In the following experiments, we first consider the scenario in which equal weights are assigned to different targets, indicating uniform attention to all the targets. Subsequently, we assign varying weights to the targets based on their properties, such as position and direction of movement, showcasing the flexibility of the proposed \gls{cdrl} framework.

\subsection{Numerical Results}
\subsubsection{Case I - Equal Weights}

\begin{figure}[h!]
    \centering
    \includegraphics[width=.4\textwidth]{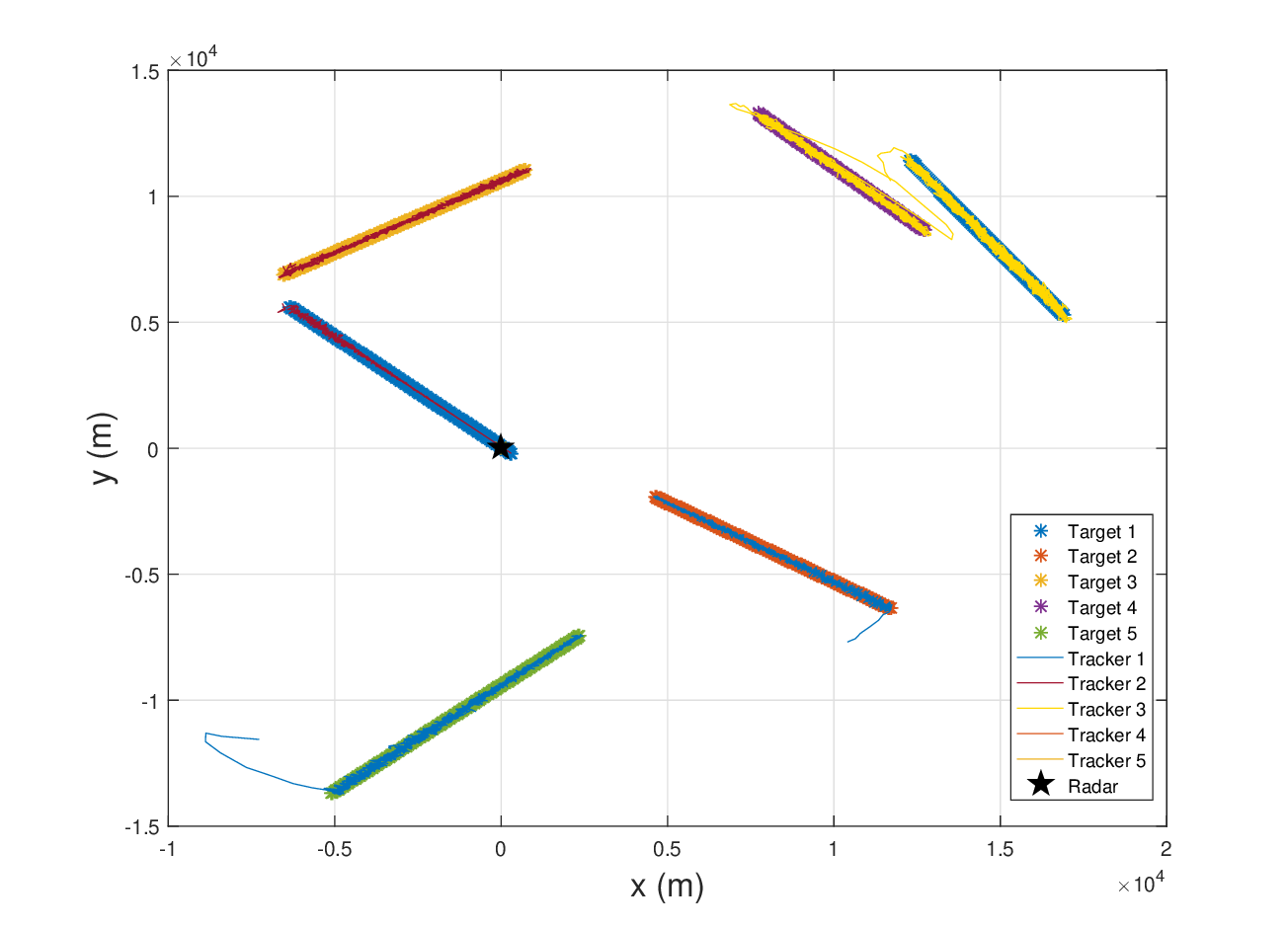}
    \caption{Case I: Trajectories of the targets}
    \label{traj}
\end{figure}

\begin{figure}[h!]
    \centering
    \includegraphics[width=.4\textwidth]{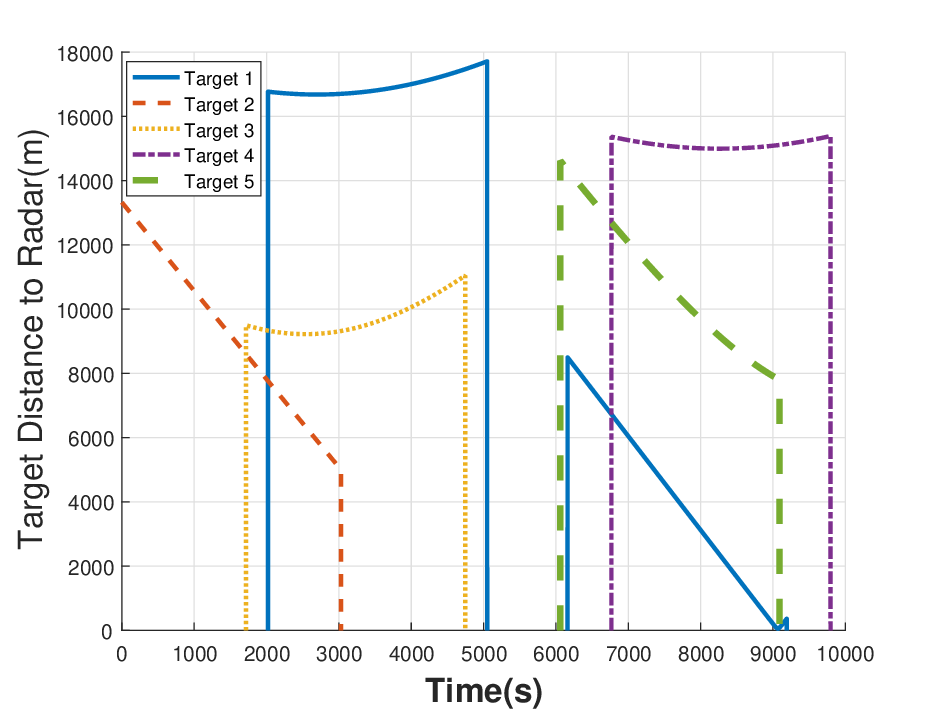}
    \caption{Case I: Targets' Distances to the Radar}
    \label{dist}
\end{figure}

\begin{figure}[h!]
    \centering
    \includegraphics[width=.4\textwidth]{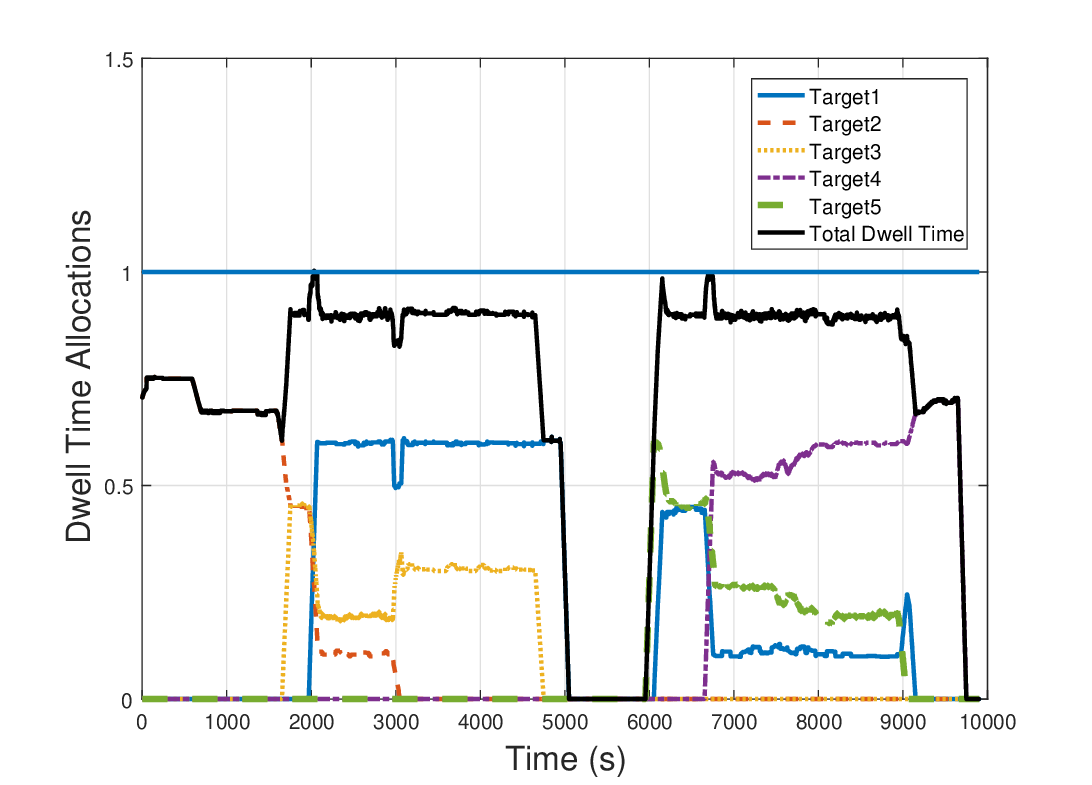}
    \caption{Case I: Dwell time allocation strategy learned by the \gls{cdrl} framework with DQL agents}
    \label{time_dql}
\end{figure}

\begin{figure}[h!]
    \centering
    \includegraphics[width=.4\textwidth]{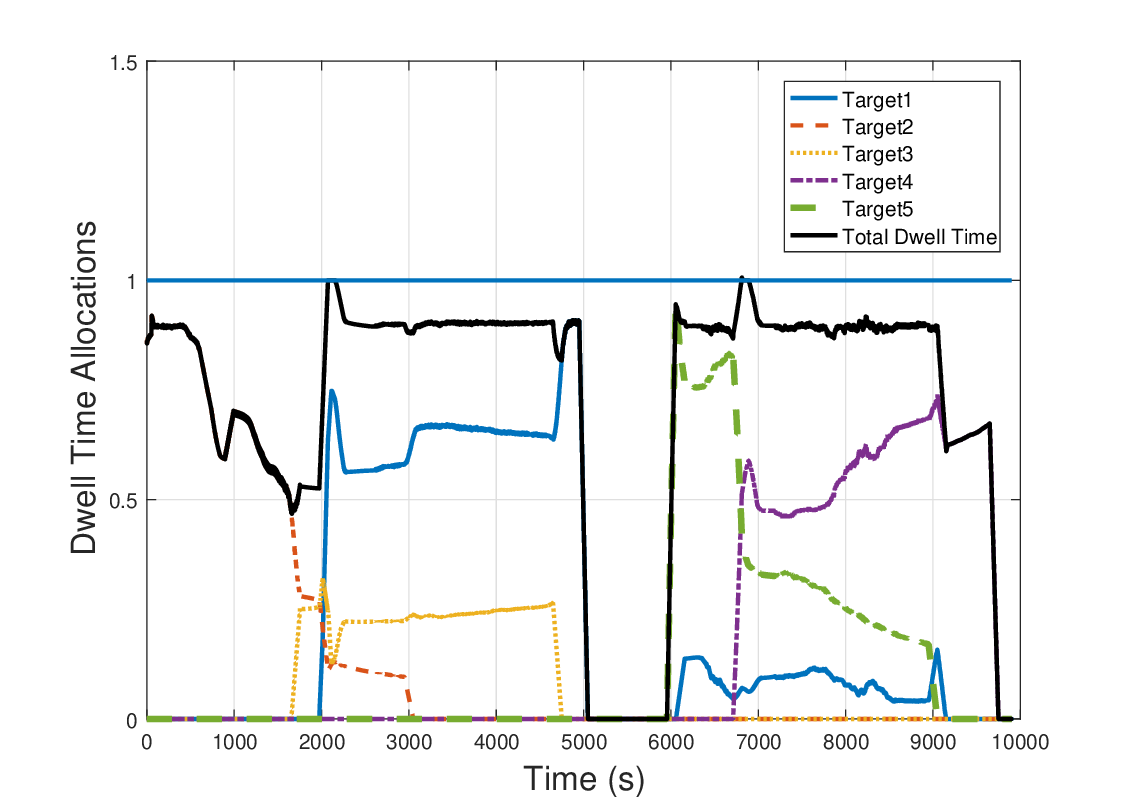}
    \caption{Case I: Dwell time allocation strategy learned by the \gls{cdrl} framework with the DDPG agent}
    \label{time_ddpg}
\end{figure}

In Case I, equal weights are assigned to tracking different targets. Fig. \ref{traj} depicts the trajectories of the targets and Fig. \ref{dist} plots the distances of the targets to the radar in the testing environment. The target spawning times and their initial positions and velocities are randomly generated. The targets are randomly generated at locations within the range of 5000m to 15000m from the radar. The magnitude of the velocity falls within the range of 0.5m/s to 2m/s. To maintain diversity in the target population and prevent tracking scenarios from becoming static, targets are automatically removed from the environment after they have existed for over 3000 time steps, regardless of their tracking status.

The objective of this work is to find an efficient time allocation strategy for tracking targets present in the environment, and to utilize the remaining time to search for and detect newly emerging targets. The strategies learned by the \gls{cdrl} framework with DQL and DDPG are presented in Fig. \ref{time_dql} and Fig. \ref{time_ddpg}, respectively. Based on these simulation results, we have the following insights regarding the radar resource management strategies  learned by the DRL agents: 

\begin{itemize}
    \item In tracking targets that are farther from the radar, more dwell time is allocated. For instance, Target 1 is the farthest from the radar during the time interval [2000, 5000] and Target 4 is the farthest during [7000, 9000]. During these intevals, radar allocates the most dwell time to tracking the corresponding farthest targets.

    \item DRL-based decision making at the radar tends to allocate more time to scanning when the demand for tracking tasks is low. This is evident during time intervals [0, 2000] and [9000, 9500], in which only one target is present in the environment. During these intervals, the total tracking time budget determined by the learning agent is kept low, allowing more time to be allocated to the scanning phase.
\end{itemize}

\begin{figure}[h!]
    \centering
    \includegraphics[width=.4\textwidth]{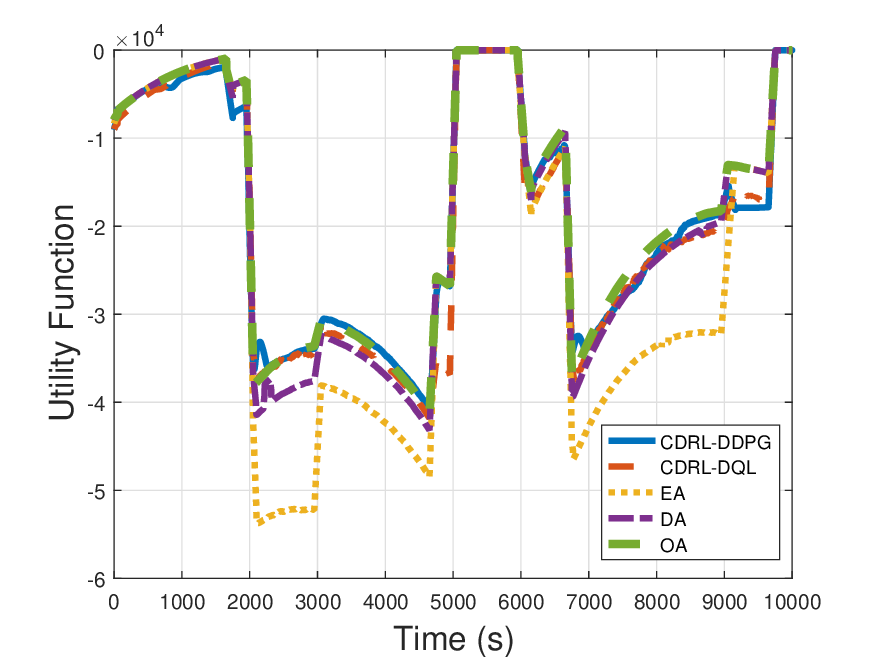}
    \caption{Case I: Comparison of utility function values}
    \label{u_comp}
\end{figure}

\begin{figure}[h!]
    \centering
    \includegraphics[width=.4\textwidth]{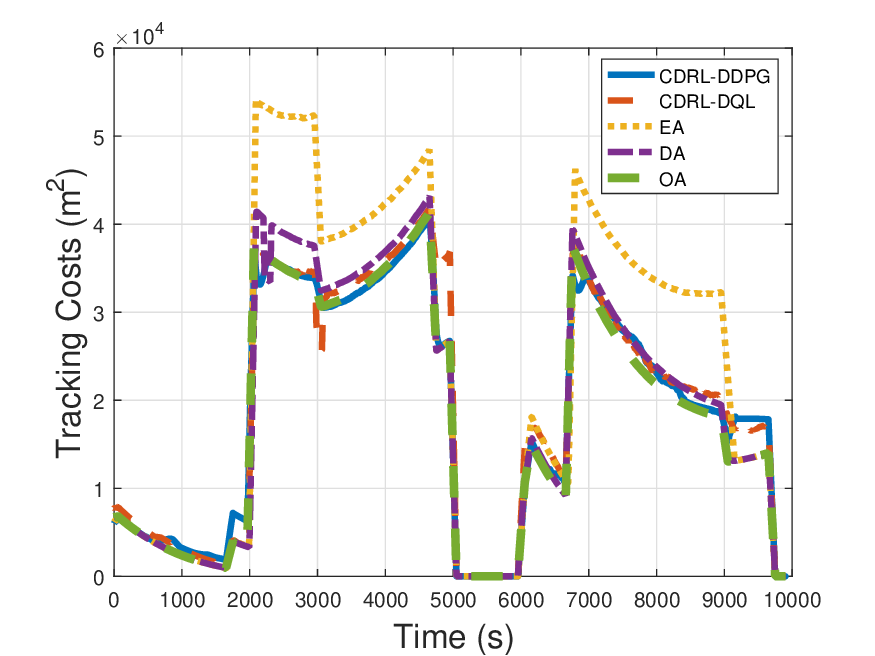}
    \caption{Case I: Comparison of tracking costs}
    \label{c_comp}
\end{figure}

We have tested the DQL and DDPG based agents in the same environment with a duration of 10000 time steps and evaluated the performance in terms of the utility function ($U_t(\{\tau_t^n\}_{n=1}^N)$), tracking costs ($\sum_{n=1}^N c_t^n$), and scanning performance (quantified via the number of missed targets $N^{miss}$ and average track initiation time $T_{init}$) as shown in Figs. \ref{u_comp}-\ref{c_comp} and Table \ref{tab:case1}. In these results, we consider the following benchmarks:

\begin{itemize}
    \item \textbf{Equal allocation (EA)}: When there exist tracked targets, ($1-\delta$) fraction of the time budget is allocated to the scanning task, while the remaining time $\delta T_0$ is evenly distributed among the tracked targets. The range of $\delta$ is $\delta\in[0, 1]$. If there is no tracked target, the entire available time will be allocated to the scanning task.

    \item \textbf{Distance-based allocation (DA)}: The radar allocates ($1-\delta$) fraction of the time to the scanning task and $\delta$ fraction of the time to the tracking tasks. When multiple targets are being tracked in the environment, the time allocated to tracking each target is proportional to its estimated distance to the radar (i.e., $\delta T_0$ is divided proportionally according to the estimated distances). Hence, targets farther away will be dedicated more dwell time compared to nearby targets. If there is a single target, the entire $\delta T_0$ is dedicated to the tracking of this target. If there is no target being tracked, all the available time $T_0$ will be allocated to scanning.

    \item \textbf{Optimization-based allocation (OA)}: The work in \cite{schope2019optimal} proposed an optimization-based approach for dwell time allocation. Under the assumptions that the targets are known already and Kalman filter is in steady state, an expression is obtained for the time-invariant Kalman filter error covariance matrix in terms of measurement and maneuverability noise statistics.    Then, optimal dwell time allocation across multiple tracking tasks is determined in this time-invariant environment using Lagrangian relaxation through an iterative procedure. More specifically, the algorithm minimizes tracking costs while adhering a total budget constraint of $\delta T_0$, with the remaining budget allocated to scanning. For each target, the approach decomposes the original problem and solves a sub-optimization problem based on measurement noise and maneuverability statistics and current tracking state to determine the optimal dwell time. The Lagrange multiplier is iteratively adjusted using the subgradient method until the budget constraint is satisfied. It is important to note that in time-varying scenarios with mobile targets, this optimization procedure needs to be repeated periodically to update dwell time allocation as conditions change.  
\end{itemize}

These three benchmarks are selected to provide a comprehensive evaluation of the proposed CDRL framework from different perspectives. Equal Allocation (EA) represents a simple baseline that requires no environmental knowledge or computation, serving as a lower bound for performance. Distance-based Allocation (DA) incorporates a single heuristic factor (target distance) that is known to influence tracking difficulty, allowing us to assess whether simple heuristics can achieve comparable performance to learning-based approaches. Optimization-based Allocation (OA) represents a principled approach that solves the tracking time allocation problem using an optimization-based approach at each time step, providing an upper bound on what can be achieved with full environmental knowledge and computational resources. Together, these benchmarks span the spectrum from simple heuristics to optimization-based methods, enabling us to evaluate both the necessity of learning-based approaches and the efficiency gains achieved by the proposed CDRL framework.

It is important to note that the benchmark methods differ in their ability to incorporate environmental knowledge. The decision-making of learning-based methods (CDRL-DQL and CDRL-DDPG) and the optimization-based allocation (OA) incorporates the real-time feedback from the environment, and they can adapt to a changing environment without prior knowledge of the changes. In contrast, the heuristic baselines—equal allocation (EA) and distance-based allocation (DA)—do not have access to such information. EA distributes resources uniformly regardless of context, while DA allocates time purely based on target distance from the radar. Neither approach can leverage knowledge about the changes in the environment.

We first set $\delta=\Theta_{max}$ for fair comparisons with the \gls{cdrl} framework, ensuring equivalent resource allocation constraints across all methods. In Figs. \ref{u_comp} and \ref{c_comp}, we observe that the two DRL-based resource management strategies (utilizing DQL and DDPG) achieve significantly higher utility function values and incur much smaller tracking costs compared to the EA scheme. This is expected since EA is highly suboptimal in the considered dynamic environment in which the number of targets and their distances to the radar vary over time. The distance-based time allocation strategy exhibits lower performance compared to the learning-based algorithms, highlighting the necessity of a learning-based approach for effective decision-making in highly dynamic environments. We also notice that optimization, DQL and DDPG based resource management policies attain similar performances. This demonstrates the efficacy of learning-based policies since optimization-based results are obtained by solving an optimization problem at each time step. Specifically, within an episode of $10000$ time steps, $10000$ optimization problems are solved. On the other hand, once the RL agents are well trained, they make decisions by executing the learned policy Here, we further note that DDPG offers significant advantages in neural network efficiency. Different from DQL, the DDPG framework supports continuous action spaces, avoids the potential exponential growth in the number of actions, and makes decisions on all dwell times simultaneously.

\begin{table}[htbp]
\caption{Case I: Comparison of Track Initiation Time}
\label{tab:case1}
\begin{tabular}{|c||c|}

  \hline
  Frameworks 
  & Average Track Initiation Time $T_{init}$ \\
  \hline
  \gls{cdrl}-DDPG 
  & 6s\\
  \hline
  \gls{cdrl}-DQL 
  & 7.83s\\
  \hline
  EA 
  & 10.50s\\
  \hline
  DA & 14.67s\\
  \hline
  OA & 20.33s\\
  \hline
  
\end{tabular} 
\end{table}

Scanning performance comparisons are presented in Table \ref{tab:case1}. We measure performance via the average track initiation time $T_{init}$, which represents the average time required for the radar to establish a target's track after it appears in the scene. In our model, track initiation occurs when a target is detected in at least $M$ out of $N$ consecutive scans, with numerical results using a $3$-of-$4$ track initiation model. Since successful detection requires repeated confirmation, track initiation time $T_{\text{init}}$ may span several seconds. In the table, we observe that the \gls{cdrl}-DDPG achieves the lowest average $T_{init}$ among all schemes, with \gls{cdrl}-DQL achieving the second-best scanning performance. The benchmark algorithms (EA, DA, and OA) fail to achieve comparable scanning performance because they allocate a fixed $(1-\delta)T_0$ amount of time for scanning when tracked targets are present. The comparison in Table \ref{tab:case1} demonstrates the importance of dynamic time allocation between scanning and tracking tasks.

\begin{figure*}[t]
    \centering
    \subfloat[Utility Function\label{u_comp_c1}]{%
        \includegraphics[width=0.32\textwidth]{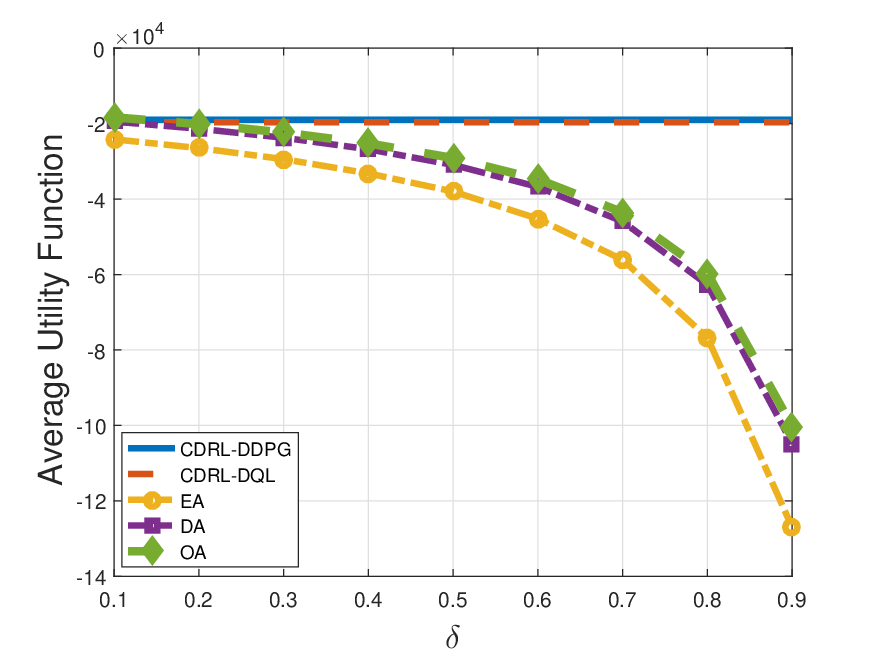}}
    \hfill
    \subfloat[Tracking Costs\label{c_comp_c1}]{%
        \includegraphics[width=0.32\textwidth]{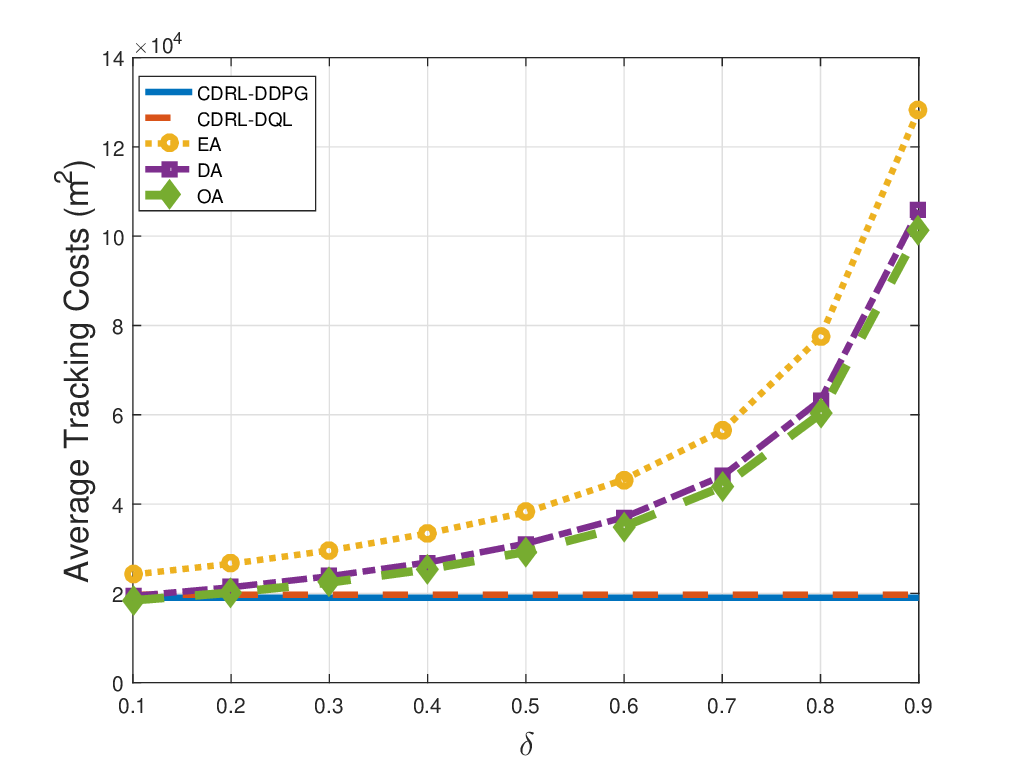}}
    \hfill
    \subfloat[Track Initiation Time\label{s_comp_c1}]{%
        \includegraphics[width=0.32\textwidth]{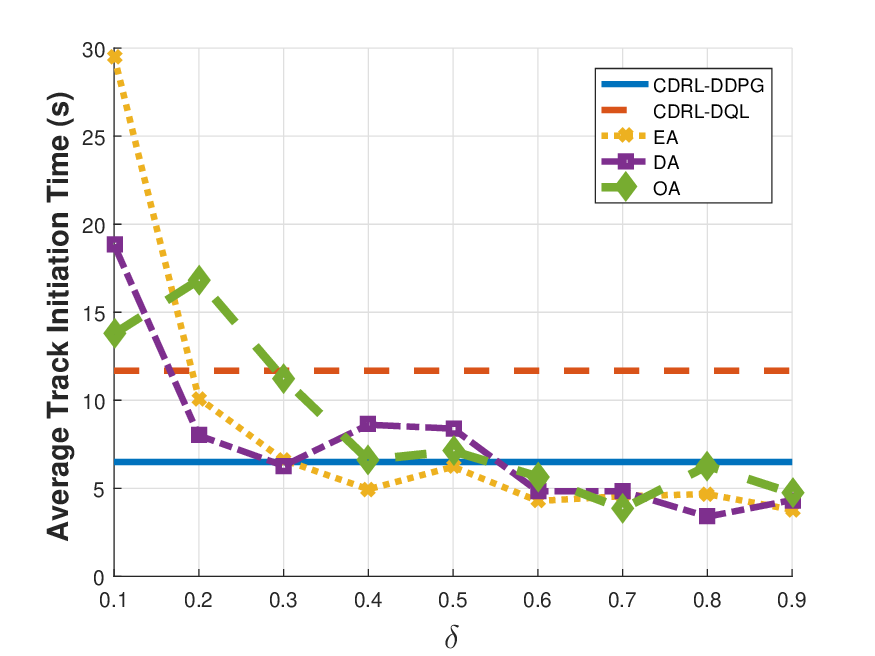}}
    \caption{Performance comparison between \gls{cdrl} and benchmark algorithms as $\delta$ varies}
    \label{perf_com_c1}
\end{figure*}

We further evaluate the benchmark algorithms (EA, DA, and OA) under varying scanning time allocation portions, $\delta$, ranging from 0.1 to 0.9. The performance metrics, averaged across multiple episodes, are presented in Fig. \ref{perf_com_c1}. The goal is to show how different scanning time allocations affect the overall performance of each benchmark algorithm.

From Fig. \ref{perf_com_c1}, we have the following conclusions:

\begin{itemize}
    \item As the fraction of time $\delta$ allocated to scanning increases, benchmark algorithms expectedly exhibit improved scanning performance but degraded tracking performance. The overall utility function tends to decrease with increasing $\delta$ as seen in Fig. \ref{perf_com_c1}(a).
    
    \item As $\delta$ approaches 0.1, benchmarks achieve utility function values and tracking costs comparable to those of the \gls{cdrl} framework, as shown in Fig. \ref{u_comp_c1} and Fig. \ref{c_comp_c1}. However, this comes at the cost of significantly increased track initiation time $T_{init}$, as evident in Fig. \ref{s_comp_c1}. This observation demonstrates the importance of dynamic time allocation between tracking and scanning tasks, a capability provided by the \gls{cdrl} framework.
    
    \item OA slightly outperforms \gls{cdrl} when $\delta=0.1$ in terms of the utility function and tracking performance as shown in Fig. \ref{u_comp_c1} and Fig. \ref{c_comp_c1}. Another advantage of the OA method is that it does not require prior training, unlike the \gls{cdrl} framework. This also makes OA appealing in scenarios where training data is scarce or offline learning is impractical. However, OA has the following weaknesses:
    \begin{itemize}
        \item [(1)] As mentioned previously, OA fails to efficiently allocate time between tracking and scanning. In particular, OA only addresses tracking time allocation. The metric of the number of undetected targets $N_t^{miss}$ (that we employ in training in the utility function and reward formulation, assuming the knowledge of ground truth) cannot be utilized in an online fashion in the optimization since this information is not available in test time. 

        \item [(2)] OA requires full knowledge of the environment parameters for decision-making, while the proposed \gls{cdrl} framework only requires the information in the agent state given in (\ref{state}).

        \item [(3)] OA exhibits significantly higher computational complexity compared to the proposed \gls{cdrl} framework. As also noted before, an optimization problem is solved iteratively at each time step in the OA method. With identical hardware configurations, \gls{cdrl} requires 1.9 milliseconds for each decision-making process while OA takes almost three times more with 5.4 milliseconds per decision.
    \end{itemize}
\end{itemize}

\subsubsection{Case II - Varying Weights}

In this section, we assume that the different tracked targets in general have different weights that can vary over time. 
The weights $\{w_t^n\}_{n=1}^N$ can depend on various properties of the targets, such as their positions or velocities. For instance, if we need to protect certain assets in the area, we should prioritize tracking targets that move closer to these assets.

\begin{figure}[h!]
    \centering
    \includegraphics[width=.4\textwidth]{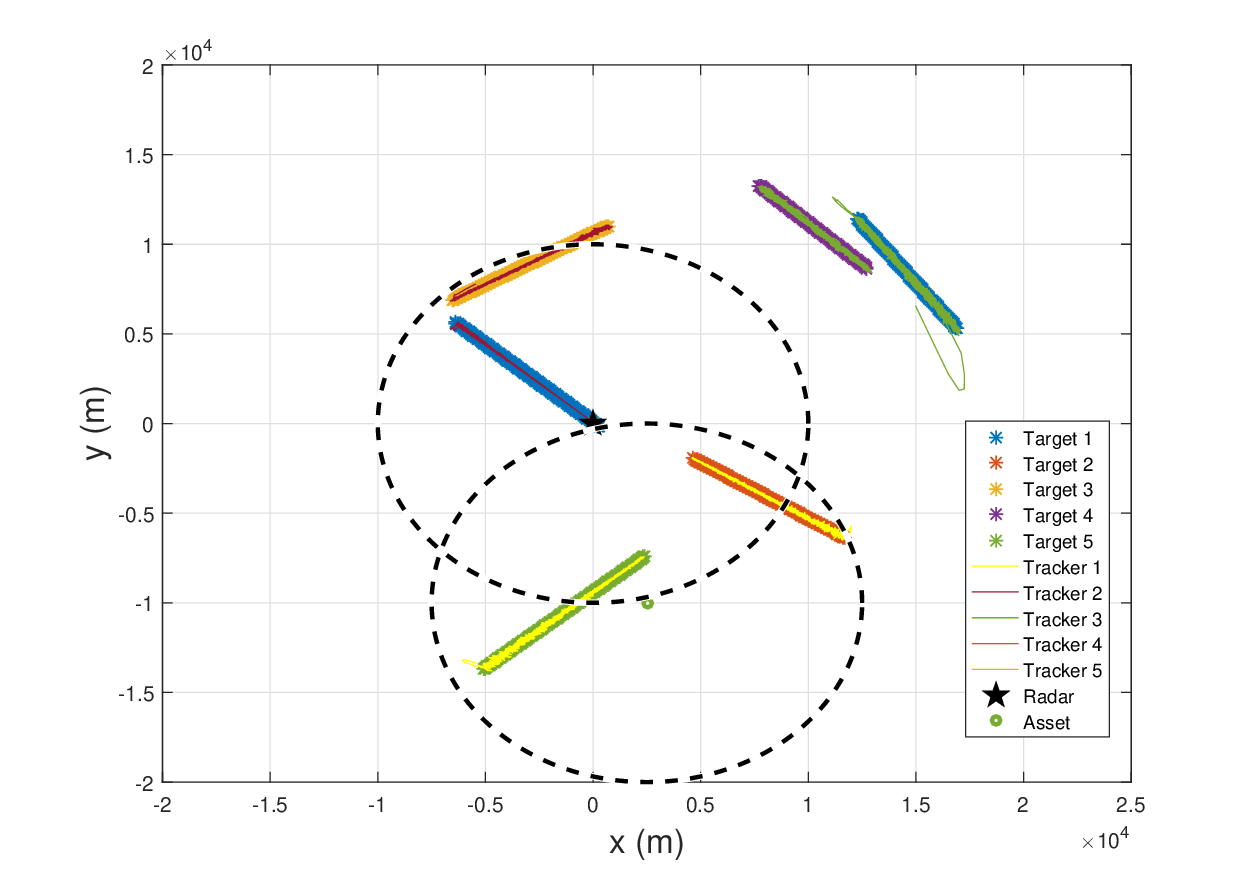}
    \caption{Case II: Trajectories of the targets}
    \label{traj2}
\end{figure}

Specifically, we assume that there is an asset located at ($2500$m, $-10000$m) as shown in Fig. \ref{traj2}. The trajectories of the targets remain the same as in Case I. When targets move too close to either the radar or the asset, it is crucial to pay greater attention to tracking these targets to ensure more accurate estimations. To achieve this objective, we design the weights  $\{w_t^n\}_{n=1}^N$ as
\begin{equation}
w_t^n = 
\begin{cases}
\min\{10, \quad 4+10000/d_t^n\}, & \text{if } d_t^n \leq 10000, \\
1 & \text{otherwise}
\end{cases}
\label{case2:weights}
\end{equation}
where \(d_t^n=\min\{d_{r,t}^n, d_{a,t}^n\}\). Here, \(d_{r,t}^n\) and \(d_{a,t}^n\) represent the current estimated distances of target \(n\) to the radar and the asset, respectively. We note that these distances are computed based on the current estimated locations of the target. The smaller of these two distances is used to determine the weight assigned to the target.

\begin{figure}[h!]
    \centering
    \includegraphics[width=.4\textwidth]{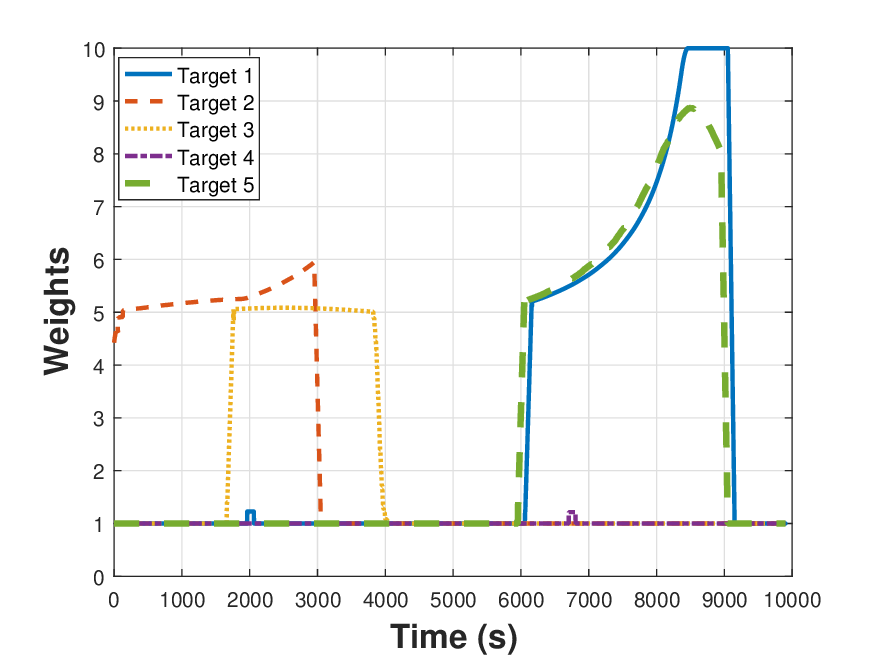}
    \caption{Case II: Weights assigned in tracking targets}
    \label{w_2}
\end{figure}

By designing the weights according to (\ref{case2:weights}), we place more emphasis on tracking the targets that move too close to either the radar or the asset. Once a target enters such a high-risk zone as shown in Fig. \ref{traj2}, the weight assigned to this target starts increasing as the target approaches further. The values of the weights are plotted in Fig. \ref{w_2}. We observe that the weights assigned to Target 1 and Target 2 increase since they approach the radar system. The weight assigned to Target 3 does not change significantly since its distance to the radar remains relatively constant, but it still maintains a certain weight because it moves within the high-risk zone near the radar. The weight assigned to Target 5 increases because it spawns within the high-risk zone around the asset and continues to approach it over time.

\begin{figure}[h!]
    \centering
    \includegraphics[width=.4\textwidth]{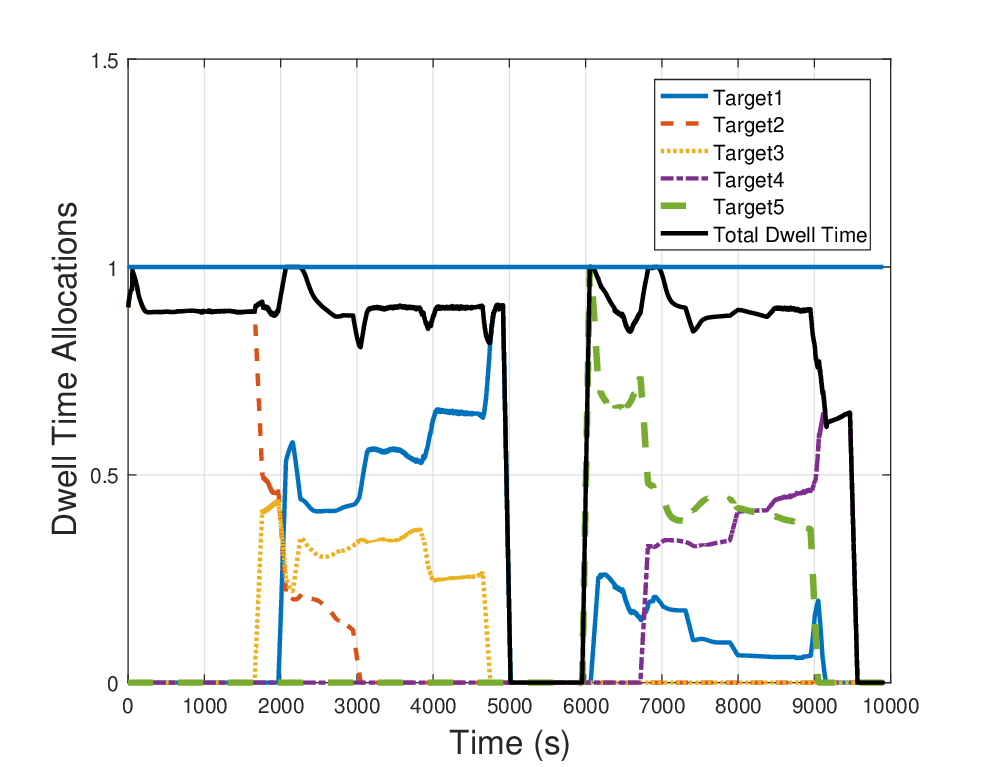}
    \caption{Case II: Dwell time allocation strategy learned by the \gls{cdrl} framework with the DDPG agent}
    \label{cas2: time_ddpg}
\end{figure}

In Case II, the time allocation strategy learned by the DDPG-based \gls{cdrl} framework is demonstrated in Fig. \ref{cas2: time_ddpg}. Compared to Fig. \ref{time_ddpg}, we have the following observations:

\begin{itemize}
    \item The radar system allocates more time to tracking Target 2 during the interval [0, 2000], to tracking Target 3 during the interval [2000, 4000], and to tracking Target 5 during the interval [6000, 9000] because the weights assigned to these targets increase during the corresponding intervals. Hence, the agent effectively learns to allocate more resources to targets that require higher levels of attention.

    \item The increase in weights leads to more time being allocated to tracking the targets, thereby reducing the time available for the scanning task, which is evident during the interval [0, 2000] and [6000, 7000].
\end{itemize}

\subsubsection{Case III - Maneuvering Targets and Beam Misalignment}

In Case II, the importance of tracking a target is addressed by the weight assigned to the target. In this section, we consider other scenarios in which greater attention is required to tracking a specific target, e.g., due to its high maneuverability. 

Heretofore, we have not accounted for potential beam misalignment between the radar and the targets. 
In practice, the radar directs a tracking beam toward the estimated location of the target. If the target is highly maneuvering, misalignment can occur between the angular direction of the radar beam and the true azimuth angle of the target, and such beam misalignment leads to a reduction in the received SNR. To address this, we modify the received SNR in (\ref{SNR}) to 
\begin{equation}
	\text{SNR}_t(\tau_t, r_t) = \text{SNR}_0\left(\frac{\tau_t}{\tau_0}\right)\left(\frac{r_t}{r_0}\right)^{-4} L_{bm}
    \label{SNR_bm}
\end{equation}
where the beam misalignment loss $L_{bm}$ is expressed as
\begin{equation}
L_{bm} = 
\begin{cases} 
  \cos^i(|\Theta - \hat{\Theta}|) & \text{if } |\Theta - \hat{\Theta}|\leq\frac{\pi}{2} \\
  0 & \text{if } |\Theta - \hat{\Theta}|>\frac{\pi}{2}.
\end{cases}.
\label{BM_c}
\end{equation}
Above, the antenna pattern of the radar beam is characterized by the exponent term $i$. In particular, a larger $i$ corresponds to a narrower beam pattern. In the numerical results, we set $i=2$. $\Theta$ and $\hat{\Theta}$ denote the true azimuth angle and estimated azimuth angle of the target, respectively.

As also noted above, highly maneuverable targets will experience greater beam misalignment losses, resulting in a lower received SNR, and consequently a higher tracking cost. The experiments in this section are designed to test whether the proposed framework learns to allocate more dwell time for tracking highly maneuverable targets.

\begin{figure}[h!]
    \centering
    \includegraphics[width=.4\textwidth]{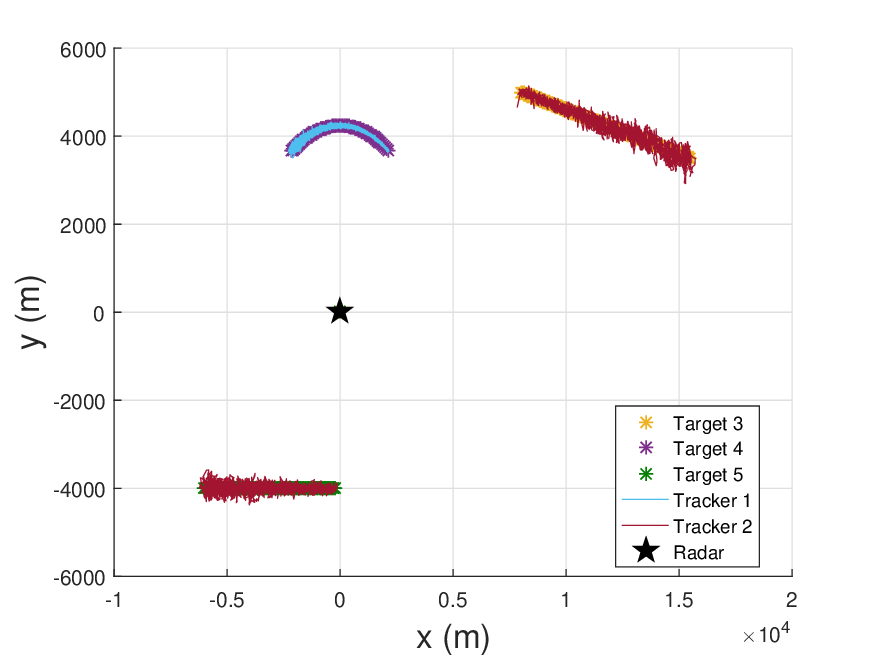}
    \caption{Case III: Trajectories of the targets}
    \label{traj3}
\end{figure}

\begin{figure}[h!]
    \centering
    \includegraphics[width=.4\textwidth]{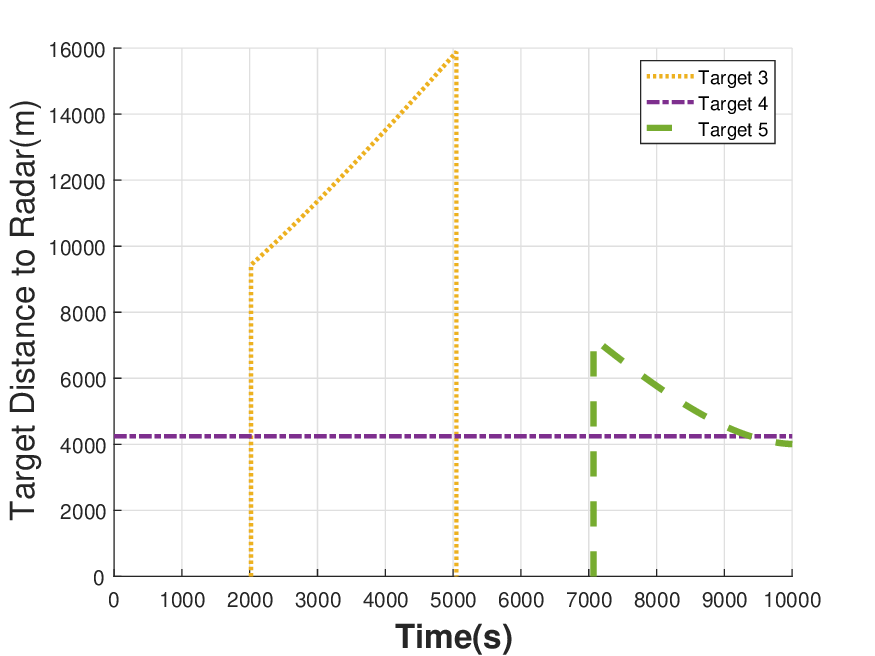}
    \caption{Case III: Targets' distances to the radar}
    \label{dist3}
\end{figure}

We assume that three targets are present in the environment. Target 3 enters the environment at $t=2000$s, with an initial location at (8000m, 5000m) and a velocity of (1m/s, -0.2m/s). Target 5 enters the environment at $t=7000$s, with the initial location at (-6000m, -4000m) and velocity of (0.8m/s, 0m/s). Target 4 is designed to follow a circular trajectory centered around the radar system, moving along this curved path back and forth at a constant angular velocity. The trajectories of the targets are shown in Fig. \ref{traj3}, and the distances to the radar of the targets are shown in Fig. \ref{dist3}. Due to its circular trajectory, Target 4 maintains a constant distance from the radar. 

\begin{figure}[h!]
    \centering
    \includegraphics[width=.4\textwidth]{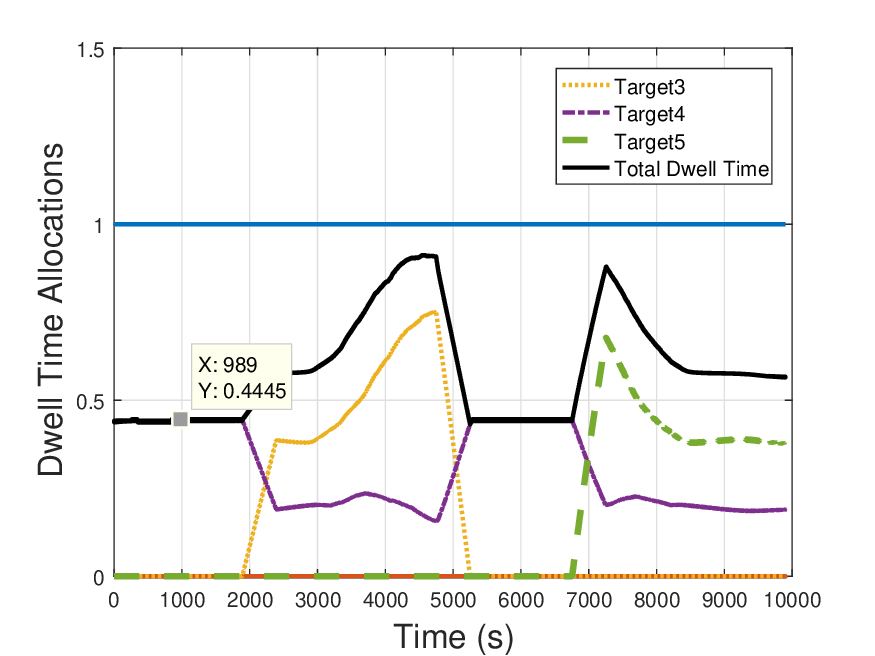}
    \caption{Case III: Dwell time allocation strategy with faster Target 4 ($\omega = 0.5$ rad/s)}
    \label{case3:fast}
\end{figure}

\begin{figure}[h!]
    \centering
    \includegraphics[width=.4\textwidth]{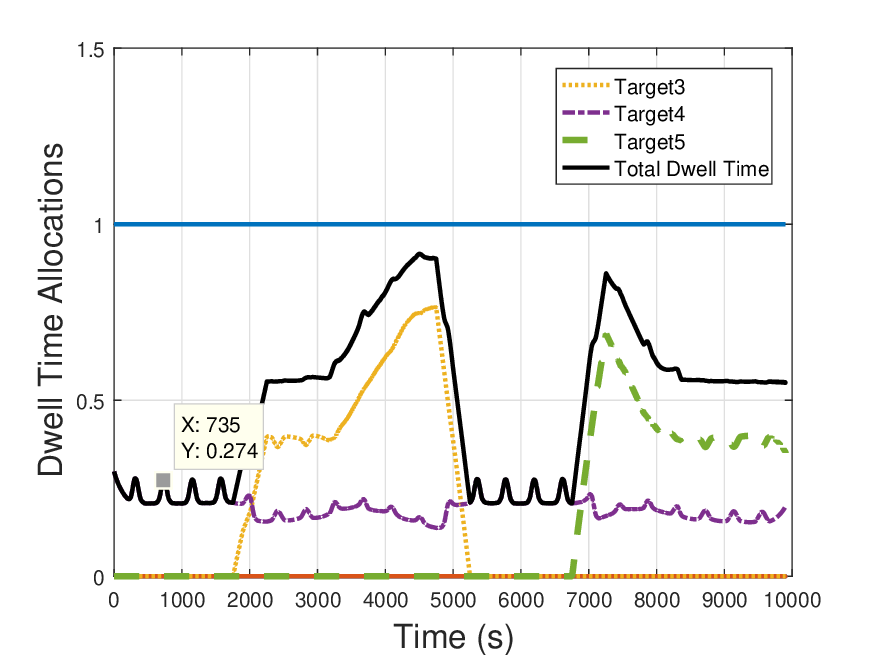}
    \caption{Case III: Dwell Time allocation strategy with slower Target 4 ($\omega = 0.005$ rad/s)}
    \label{case3:slow}
\end{figure}

We tested the proposed \gls{cdrl}-DDPG framework in two scenarios: a slower Target 4 with an angular velocity of $\omega = 0.005$ rad/s and a faster Target 4 with an angular velocity of $\omega = 0.5$ rad/s. The time allocation strategies learned in these two scenarios are depicted in Fig. \ref{case3:fast} and Fig. \ref{case3:slow}, respectively. Fig. \ref{case3:fast} displays the time allocation strategy learned in the presence of the faster Target 4. Compared to the strategy with the slower Target 4 as shown in Fig. \ref{case3:slow}, the proposed framework learns to allocate more time for tracking Target 4 with the higher angular velocity. This is evident during the intervals [0, 2000] and [5000, 7000]. Therefore, we conclude that the proposed \gls{cdrl} framework can recognize the maneuverability of the target and adapt its time allocation strategy accordingly in different scenarios.

The oscillating pattern in dwell time allocation for Target 4 in Fig. \ref{case3:slow} reflects the periodic variations in tracking difficulty as the target moves along its circular trajectory, resulting in  varying angular positions and, consequently, changing beam pointing requirements and associated beam misalignment effects. The faster target in Fig. \ref{case3:fast} exhibits a flatter profile due to the stronger moving average smoothing effect on higher-frequency allocation adjustments.

\subsubsection{Case IV - Unexpected Environmental Changes}

In Case IV, it is assumed that multiple high-interference regions are randomly located on the map, where the measurement noise variance for targets can significantly increase due to factors such as poor weather conditions, environmental clutter, or electromagnetic interference.

\begin{figure}
    \centering
    \includegraphics[width=0.4\textwidth]{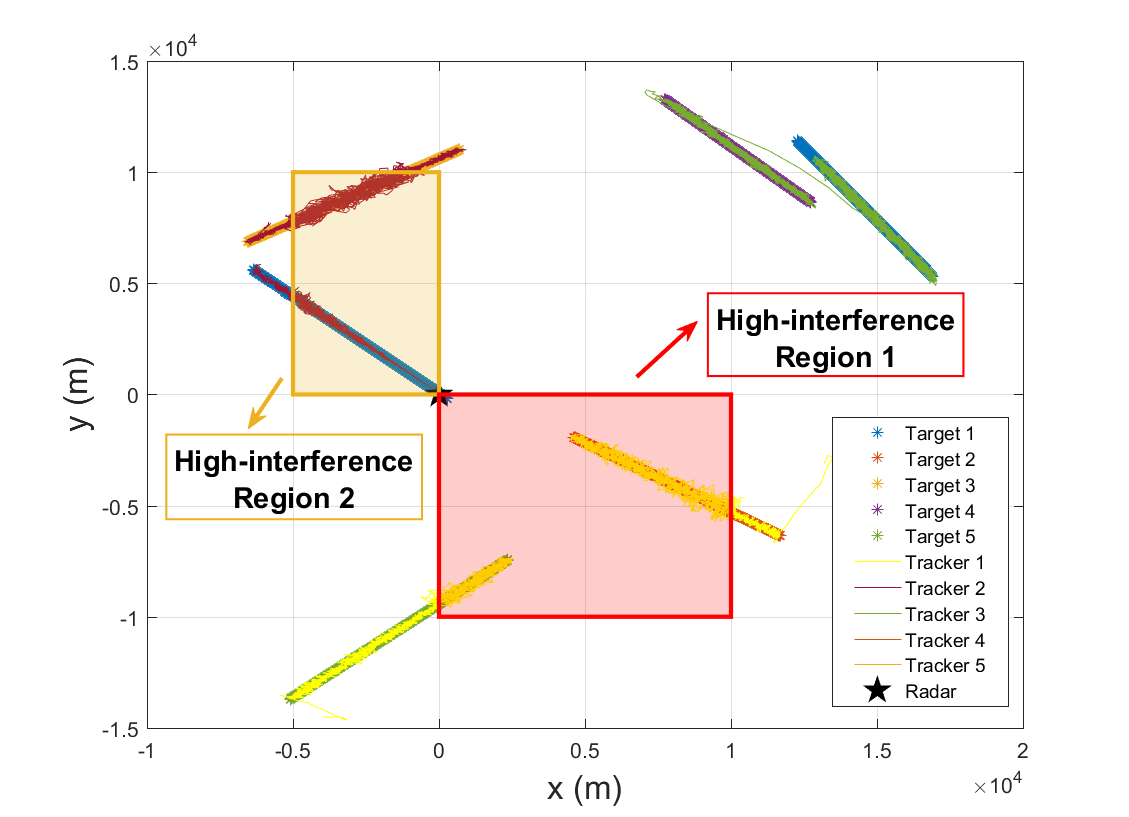}
    \caption{Case IV: Trajectories of the targets}
    \label{hazard_traj}
\end{figure}

For this case study, we define two high-interference regions where the measurement noise significantly increases due to environmental degradation such as clutter or interference, as shown in Fig. \ref{hazard_traj}. Specifically, the first high-interference region is defined within $x \in [0\,\text{m},\,10000\,\text{m}]$ and $y \in [-10000\,\text{m},\,0\,\text{m}]$, where the measurement noise variance is increased by a factor of 7. The second high-interference region spans $x \in [-5000\,\text{m},\,0\,\text{m}]$ and $y \in [0\,\text{m},\,10000\,\text{m}]$, where the noise variance is increased by a factor of 10. 

\begin{figure}[h!]
    \centering
    \includegraphics[width=.4\textwidth]{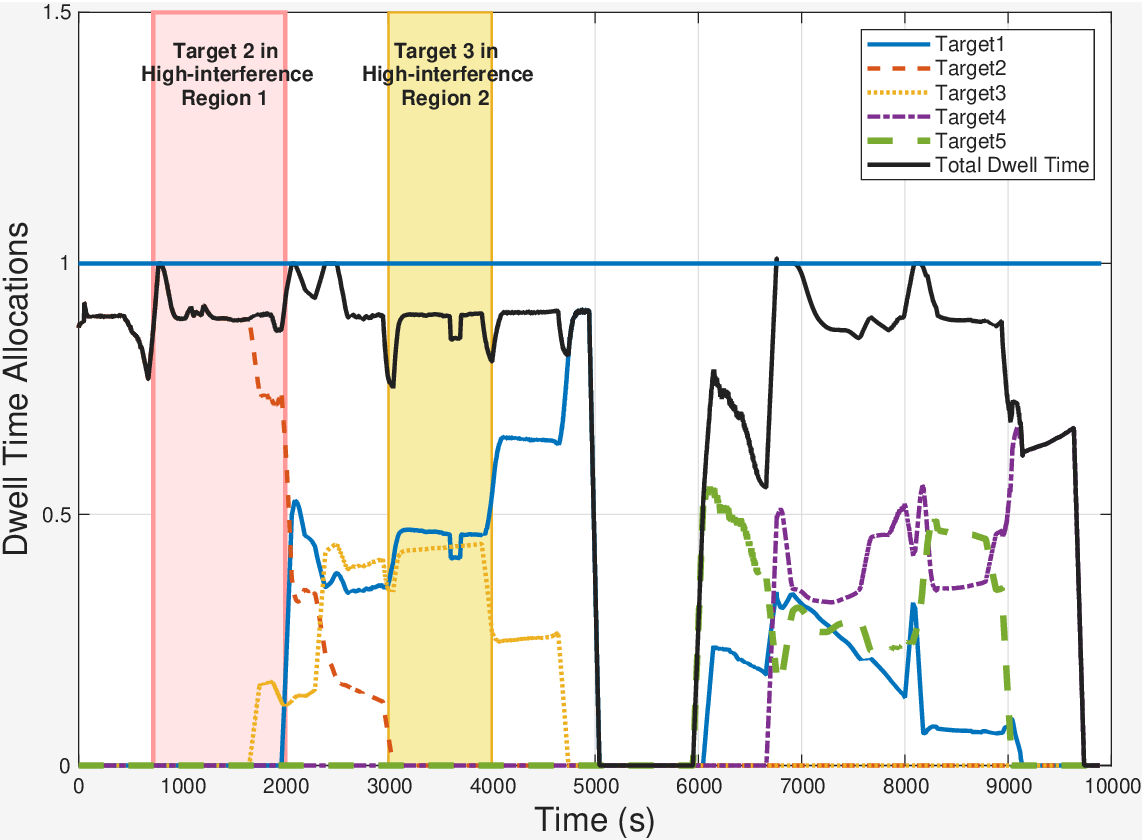}
    \caption{Case IV: Dwell time allocation strategy with high-interference region}
    \label{case4}
\end{figure}

The CDRL agent is not directly informed of the existence of these regions but is able to sense environmental changes through rising tracking costs. The adaptive time allocation strategy learned by the CDRL agent in response to this dynamic environment is illustrated in Fig.~\ref{case4}. The red and yellow shaded regions highlight time periods when Target 2 and Target 3 are within high-interference regions 1 and 2, respectively. As the measurement noise variance increases, the radar receives noisier observations, resulting in higher tracking costs as estimated by the EKF. Compared to Fig. \ref{time_ddpg} in Case I, where no high-interference region is present, the CDRL agent clearly detects the degradation in sensing quality and responds by allocating more tracking time to the affected targets during the highlighted time periods (red shaded region for Target 2, yellow shaded region for Target 3).

\begin{table*}[htbp]
\centering

\begin{tabular}{|l|c|c|c|c|}
\hline
 & \textbf{Case I} & \textbf{Case II} & \textbf{Case III} & \textbf{Case IV} \\
\hline
\textbf{\gls{cdrl}-DDPG} & $-1.90\times 10^4$ & $\mathbf{-3.54\times 10^4}$ & $-8.14\times 10^3$ & $-5.11\times 10^4$ \\
\hline
\textbf{\gls{cdrl}-DQL} & $-1.96\times 10^4$ & $-3.59\times 10^4$ & $-7.91\times 10^3$ &  $-5.06\times 10^4$ \\
\hline
\textbf{EA} & $-2.42\times 10^4$ & $-4.10\times 10^4$ & $-13.7\times 10^3$ & $-5.57\times 10^4$ \\
\hline
\textbf{DA} & $-1.94\times 10^4$ & $-3.90\times 10^4$ & $-7.96\times 10^3$ &  $-5.59\times 10^4$ \\
\hline
\textbf{OA} & $\mathbf{-1.82\times 10^4}$ & $-3.59\times 10^4$ & $\mathbf{-6.72\times 10^3}$ &  $\mathbf{-5.05\times 10^4}$ \\
\hline
\end{tabular}
\caption{Comparison of Achieved Utility Function Values}
\label{tab:utility-comparison}
\end{table*}

We summarize the performance of the proposed \gls{cdrl} frameworks as well as the EA, DA, and OA approaches in terms of the utility function in Table~\ref{tab:utility-comparison}. Our analysis yields the following observations:
\begin{itemize}
    \item OA consistently achieves near-optimal performance across all three cases. However, as noted in Case I, OA exhibits significant limitations: inability to balance time allocation between scanning and tracking, requirement of complete environmental knowledge, and high computational complexity.
    
    \item \gls{cdrl}-DDPG and \gls{cdrl}-DQL demonstrate comparable performance in all test cases. \gls{cdrl}-DDPG offers the additional advantage of preventing action space explosion by employing continuous action via the DDPG algorithm.
    
    \item EA consistently delivers the poorest performance across all scenarios. While DA performs comparably in Cases I and III, its utility function value is reduced in Cases II and IV. This suggests that although DA can offer reasonable performance in scenarios where distance is a reliable proxy for tracking importance, it lacks the adaptability to handle more complex conditions, such as varying tracking priorities or environmental changes.

\end{itemize}

\section{Conclusions}

In this paper, we have proposed a \gls{cdrl} framework to address the time allocation problem for a cognitive radar system operating in a track-while-scan mode. The radar system performs two tasks: scanning the environment for potential targets and tracking previously detected targets. A utility function was proposed to effectively balance the trade-off between scanning and tracking targets located at different geographic positions relative to the radar. We detailed the tracking and scanning model for the considered problem and formulated the time allocation problem as a constrained sequential optimization problem. We then proposed a \gls{cdrl} framework where the parameters of the neural networks and the dual variable are learned simultaneously. We designed both DQL and DDPG based \gls{drl} agents for decision-making. 

Via numerical results, we demonstrated the effectiveness of the proposed framework in jointly optimizing the scanning and tracking performance. In particular, by maximizing the utility function, the DRL agents intelligently allocate more time to tracking targets with higher tracking costs while ensuring sufficient time for the scanning task, all within budget constraints.  With uniform tracking weights, the agents learn to allocate more resources to tracking targets that are farther away. Our numerical results show that the proposed \gls{cdrl} framework consistently outperforms fixed-allocation or heuristic strategies such as Equal Allocation (EA) and Distance-based Allocation (DA) in both scanning and tracking performance. Unlike fixed heuristics, the learned policy adapts to real-time factors such as target location, motion, and priority, resulting in improved responsiveness and better use of limited resources. Although OA (optimization-based allocation) can provide strong performance under ideal conditions, it requires full environment knowledge, incurs higher computational complexity, and addresses only tracking time allocation. \gls{cdrl} achieves comparable performance with significantly more flexibility and lower overhead.

This work can be extended in several promising directions. First, the framework can be enhanced to optimize multiple radar parameters simultaneously, including waveform selection, beamform design, and transmit power allocation. Second, the system can be evaluated in more realistic environments that incorporate challenging factors like clutter and jamming, or integrated into emerging paradigms such as integrated sensing and communication (ISAC) in next-generation wireless systems. Third, while this work employs a centralized DDPG agent to avoid exponential growth in the action space, transitioning to a distributed multi-agent DRL architecture could offer further advantages. If the action space complexity can be effectively managed, a distributed approach would not only better protect user privacy but also enable more autonomous decision-making by individual agents—a critical capability for operating in challenging environments with larger numbers of targets. Fourth, the current work assumes that data association can reliably separate measurements from different targets. Future research should investigate scenarios where targets are in close proximity or their tracks cross, requiring more sophisticated data association methods such as Joint Probabilistic Data Association (JPDA) or Multiple Hypothesis Tracking (MHT) integrated with the CDRL framework to handle measurement ambiguity and prevent track swapping.

\bibliographystyle{IEEEtran}
\bibliography{ref}

\end{document}